\documentclass[aps,prb,manuscript]{revtex4-1}
\usepackage{graphicx}

\begin{document}
\draft
\title{Role of electron back action on photons in hybridizing double-layer graphene 
plasmons with localized photons}

\author{Danhong Huang$^{1}$, Andrii Iurov$^{2}$ and Godfrey Gumbs$^{3}$}

\affiliation{$^{1}$Air Force Research Laboratory, Space Vehicles Directorate, 
Kirtland Air Force Base, New Mexico 87117, USA\\
$^{2}$Center for High Technology Materials, University of New Mexico, 
1313 Goddard SE, Albuquerque, New Mexico, 87106, USA\\
$^{3}$Department of Physics and Astronomy, Hunter College of the City 
University of New York, 695 Park Avenue New York, New York 10065, USA}

\date{\today}

\begin{abstract}
Induced polarization by Dirac electrons in double-layer graphene can affect 
hybridization of radiative and evanescent fields. Electron back action 
appears as a localized optical field to modify an incident surface-plasmon-polariton 
(SPP) evanescent field. This leads to high sensitivity (beyond the diffraction 
limit) to local environments and provides a scrutiny tool for molecules or protein 
selectively bounded with carbon. A scattering matrix with frequencies around the 
surface-plasmon (SP) resonance supports this scrutiny tool and exhibits sensibly the 
increase, decrease and even a full suppression of the polarization field in the    
vicinity of a conducting surface for longer SPP wavelengthes. Moreover, triply-hybridized 
absorption peaks associated with SP, acoustic- and optical-like graphene plasmons 
become significant only at high SP frequencies, but are overshadowed by a round SPP 
peak for low SP frequencies. These resonant features (different from 3D photonic 
lattices) facilitate the polariton-only excitations, giving rise to possible 
polariton condensation for a threshold-free laser.
The current graphene-plasmon hybridization formalism can be easily generalized to other two-dimensional materials, such as silicene, germanene, molybdenum disulfide, etc. 
\end{abstract}
\pacs{PACS:}
\maketitle

\section{Introduction}
\label{sec-1}

When light is incident on a semiconductor, its energetic photons can excite electrons 
from a lower valence band to a higher conduction band, thereby creating many electron-hole pairs 
in the system\,\cite{add8,add26,ref1}.  Simultaneously, its electric-field component 
is able to push away the negatively (positively) charged electrons (holes) in opposite 
directions.  In this case, the excited electrons and holes will also exert a 
back action on the incident light, resulting from the induced optical polarization 
as a collection of local dipole moments from many displaced electrons and 
holes\,\cite{add9,add29,new}. This polarization field can further scatter incident 
photons resonantly\,\cite{ref1,add12,swk}. Therefore, the quantum nature of Dirac 
electrons\,\cite{add1,add2,add3,add4,special,chapter} will be revealed in this electron 
back action on the incident light.
\medskip

\begin{figure}
\centering
\includegraphics[width=0.48\textwidth]{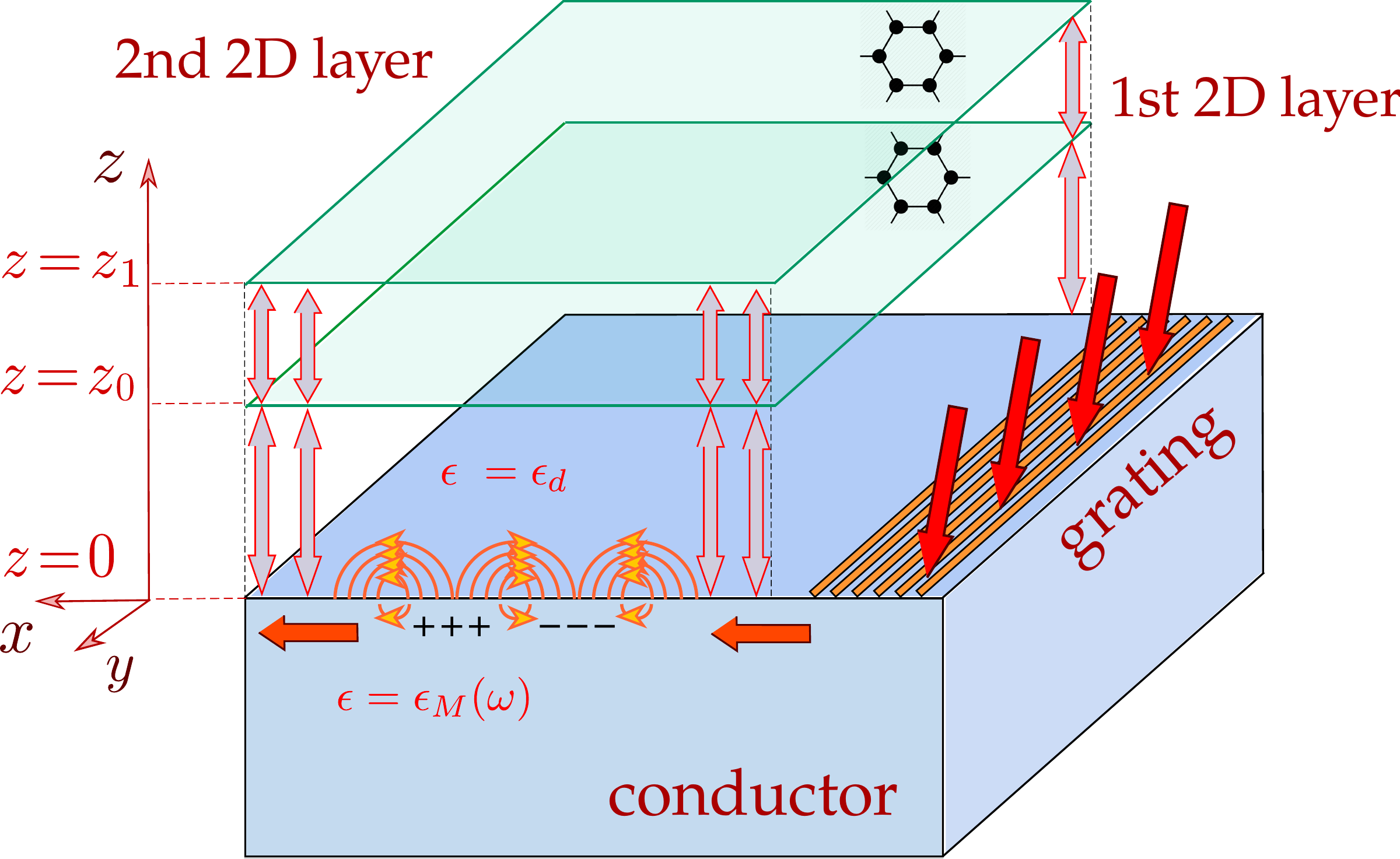}
\caption{(Color online) Schematics for a thick (semi-infinite) conductor in the region 
$z < 0$ and having a frequency-dependent local dielectric function $\epsilon_M (\omega)$. 
Two graphene layers at $ z = z_0$ and $z = z_1$ above the  surface of the conductor are embedded 
within a semi-infinite dielectric having a dielectric constant $\epsilon_d$ in the 
region $z>0$. These two-dimensional sheets are coupled to each other and also to 
the semi-infinite conductor by an electromagnetic interaction.}
\label{f1}
\end{figure}
\begin{figure}
\centering
\includegraphics[width=0.4\textwidth]{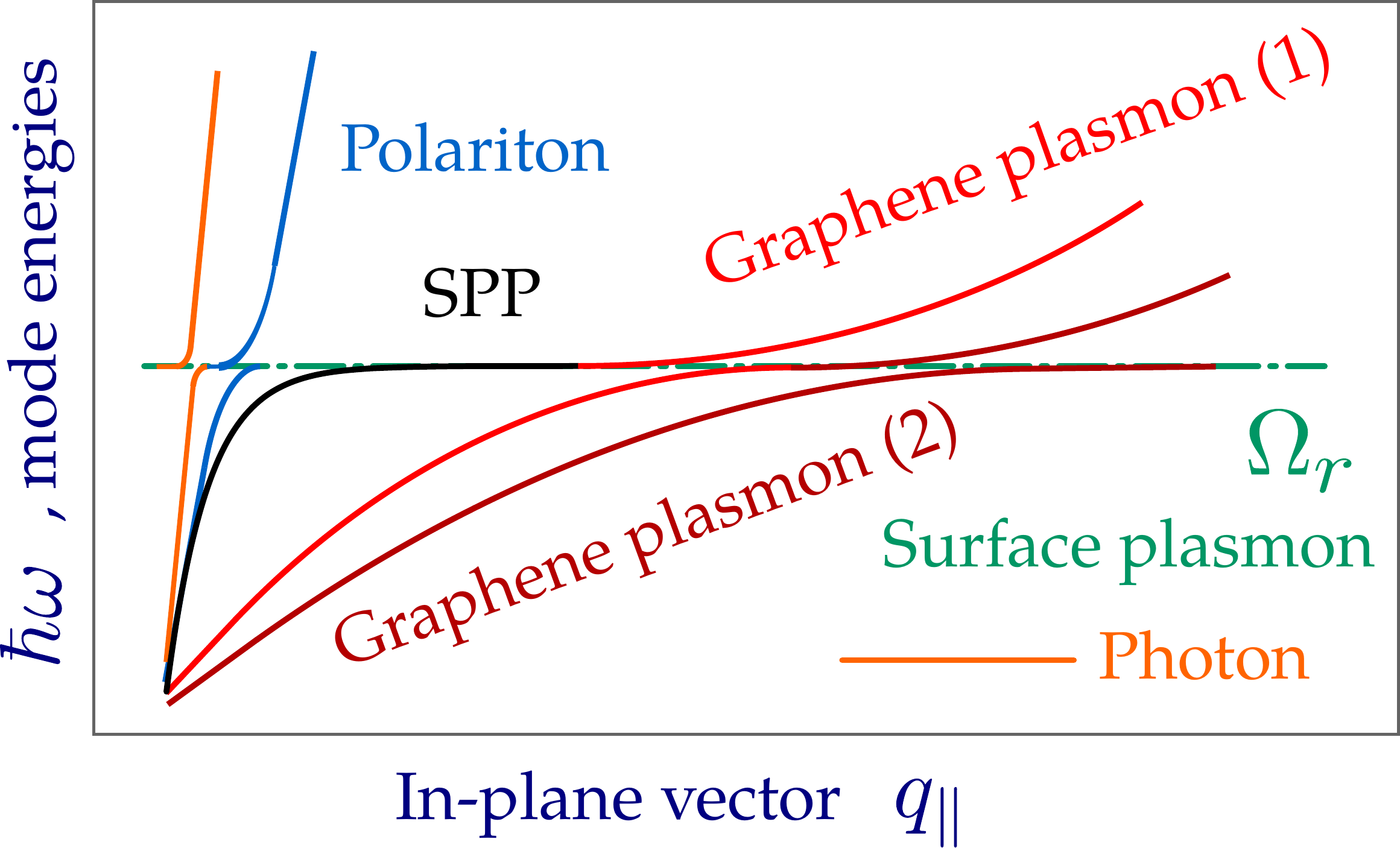}
\caption{(Color online) Schematic representation of the energy dispersion relations 
for radiative and evanescent light as well as field modes for the hybrid structure depicted 
in Fig.\,\ref{f1}. These include photons, polaritons, surface-plasmon polaritons (SPPs), 
two non-degenerate graphene plasmons (G-Ps), and surface plasmons (SPs).}
\label{f2}
\end{figure}

For the hybrid system shown schematically in Fig.\,\ref{f1}, we are faced with both radiative field 
modes, such as photons and polaritons\,\cite{oe1,oe2,oe3,oe4,add28}, and evanescent field 
modes, e.g., surface and graphene plasmons\,\cite{ritchie,add5,add6,add7}. Research on  the
optical response of graphene electrons has been previously reported\,\cite{new,add7,add10}, but 
most of those studies have been concerned with the effect due to  radiation or grating-deflection 
field coupling. In contrast to the plane-wave-like light field, we examine the role of coupling by a 
surface-plasmon-polariton (SPP) near field\,\cite{add11,add15,add27} to graphene 
electrons with a different dispersion relation from the usual linear one, i.e., 
$\omega=qc$, for light in free-space. In this paper, double graphene layers are 
placed very close to the surface of a conducting substrate so that radiative and 
evanescent fields are hybridized effectively\,\cite{add24,pan2}. Consequently, the 
non-dispersive surface-plasmon (SP) mode can hybridize successfully with radiative 
photon and polariton modes\,\cite{oe1,oe2}, as well as with the spatially-localized 
graphene plasmon (G-P) mode\,\cite{pan2,add14}, as illustrated in Fig.\,\ref{f2}.
This is quite different from three-dimensional photonic lattices\,\cite{shawn,shawn2}, 
where electrons interact with quantized multi-subband photons in the first 
Brillouin zone. 
\medskip

Such a unique dispersion relation of the hybrid light-plasmon modes should be verified experimentally 
by optical spectroscopy\,\cite{add18,add25,add31,add32}. The effective scattering 
matrix\,\cite{new,add19,add30} for such a coupled system predicts distinctive features 
neither from the graphene sheets nor from the conductor on their own, and it retains the properties 
of a longitudinal electromagnetic interaction\,\cite{ref1,new,add12} between electrons 
in double-layer graphene and a conductor. This scattering matrix can also be employed for 
constructing an effective-medium theory\,\cite{add20,add21,add22,add23} used for investigating 
the optical properties of inserted biomolecules and metamaterials between the graphene 
sheets and the surface of the conductor. As a whole, a local environmentally sensitive 
super-resolution near-field imaging\,\cite{xzhang} (beyond the diffraction limit)
should be possible for functionalized biomolecules bounded with either metallic 
nanodots and nanorods or carbon atoms of graphene\,\cite{add13,ieee}.
\medskip
 
The remainder of the paper is arranged as follows. In Sec.\,\ref{sec-2}, we present the 
Green's function formalism for the hybridized system depicted in Fig.\,\ref{f1}. By using 
Green's functions, an integral equation is established from Maxwell's equations 
by treating double-layer graphene as a localized polarization-field source. In 
Sec.\,\ref{sec-3}, the optical-response function of Dirac electrons in gapped
graphene is obtained after explicitly calculating the density-density correlation 
function at low temperatures. Based on linear-response theory for the localized 
graphene polarization field, we present in Sec.\,\ref{sec-4} a self-consistent 
equation for the total electric field after combining the integral equation
derived in Sec.\,\ref{sec-2} with the optical-response function calculated in 
Sec.\,\ref{sec-3}, from which a dispersion relation is obtained for the hybridized 
G-P and SP modes. In addition, a local effective scattering matrix, as well as a 
spatial distribution for the scattering field, are derived in Sec.\,\ref{sec-4}, 
which are further accompanied by an optical absorption spectrum calculated for 
hybridized G-P and SP modes. Finally, conclusions and some remarks are given in 
Sec.\,\ref{sec-5}.

\section{Green's Function for Hybridized Systems}
\label{sec-2}

Considering steady states, then from one of  Maxwell's equations we can write down the following 
equation\,\cite{ref1,new,add11} for two semi-infinite non-magnetic media 
in position-frequency space as 

\begin{equation}
\mbox{\boldmath$\nabla$$\times$\boldmath$\nabla$$\times$\boldmath$E$}({\mbox{\boldmath$r$}};\omega)-\epsilon_{b}(x_3;\omega)\,\frac{\omega^2}{c^2}\,\mbox{\boldmath$E$}({\mbox{\boldmath$r$}};\omega)
=\frac{\omega^2}{\epsilon_0c^2}\,\mbox{\boldmath${\cal P}$}^{\rm loc}({\mbox{\boldmath$r$}};\omega)\ ,
\label{e2.1}
\end{equation}
where $\mbox{\boldmath$E$}({\mbox{\boldmath$r$}};\omega)$ represents the electric component 
of an electromagnetic field, 
${\mbox{\boldmath$r$}}=({\mbox{\boldmath$r$}}_\|,x_3)=(x_1,x_2,x_3)$ is a three-dimensional 
position vector, $\omega$ is the angular frequency of the incident light. In addition,
$\mbox{\boldmath$H$}({\mbox{\boldmath$r$}};\omega)=\left(1/i\omega\mu_0\right)\mbox{\boldmath$\nabla$}\times\mbox{\boldmath$E$}({\mbox{\boldmath$r$}};\omega)$ represents the magnetic component 
of the electromagnetic field, $\epsilon_0$, $\mu_0$ and $c$ are the permittivity, permeability 
and speed of light in vacuum, respectively. Furthermore, 
$\mbox{\boldmath${\cal P}$}^{\rm loc}({\mbox{\boldmath$r$}};\omega)$ is a local polarization 
field produced by optical transitions of electrons in graphene sheets embedded in a semi-infinite dielectric, as schematically shown in Fig.\,\ref{f1}, and the position-dependent dielectric 
function takes the form 

\begin{equation}
\epsilon_{b}(x_3;\omega)=\left\{\begin{array}{ll}
\epsilon_{d}\ , & \mbox{for $x_3>0$}\\
\epsilon_{M}(\omega)\ , & \mbox{for $x_3<0$}
\end{array}\right.\ .
\label{e2.2}
\end{equation}
In Eq.\,(\ref{e2.2}), $\epsilon_{d}$ characterizes the semi-infinite dielectric 
material in the region $x_3>0$, while $\epsilon_M (\omega)=\epsilon_s-\Omega_p^2/[\omega(\omega+i0^+)]$ 
is the dielectric function of the semi-infinite conducting material in the region $x_3<0$.
For the Maxwell equation in Eq.\,(\ref{e2.1}), we have introduced the corresponding Green's 
function ${\cal G}_{\mu\nu}({\mbox{\boldmath$r$}},{\mbox{\boldmath$r$}}^\prime;\omega)$ that 
satisfies the equation\,\cite{add11}

\begin{equation}
\sum\limits_{\mu}\left[\epsilon_{b}(x_3;\omega)\,\frac{\omega^2}{c^2}\,\delta_{\lambda\mu}-\frac{\partial^2}{\partial x_{\lambda}\partial x_{\mu}}+\delta_{\lambda\mu}\,\nabla_{\bf r}^2\right]{\cal G}_{\mu\nu}({\mbox{\boldmath$r$}},{\mbox{\boldmath$r$}}^\prime;\omega)=\delta_{\lambda\nu}\,\delta({\mbox{\boldmath$r$}}-{\mbox{\boldmath$r$}}^\prime)\ ,
\label{e2.3}
\end{equation}
where $\nabla_{\bf r}^2=\sum\limits_{\mu}\,\partial^2/\partial x^2_{\mu}$ is the Laplace 
operator, $\delta_{\lambda\mu}$ is the  Kronecker delta, and the indices $\lambda,\,\mu=1,\,2,\,3$ 
specify three spatial directions. Using the Green's function determined by Eq.\,(\ref{e2.3}),
we can rewrite the Maxwell equation in Eq.\,(\ref{e2.1}) in integral form\,\cite{ref1,new}

\begin{equation}
E_{\mu}({\mbox{\boldmath$r$}};\omega)=E^{\rm inc}_{\mu}({\mbox{\boldmath$r$}};\omega)+\frac{\omega^2}{\epsilon_0c^2}\sum\limits_{\nu}\int d^3{\mbox{\boldmath$r$}}^\prime\,
{\cal G}_{\mu\nu}({\mbox{\boldmath$r$}},{\mbox{\boldmath$r$}}^\prime;\omega)\,
{\cal P}_{\nu}^{\rm loc}({\mbox{\boldmath$r$}}^\prime;\omega)\ ,
\label{e2.4}
\end{equation}
where $E^{\rm inc}_{\mu}({\bf r};\omega)$ stands for a solution for the following homogeneous 
equation\,\cite{add11}

\begin{equation}
\sum\limits_{\nu}\left[\epsilon_{b}(x_3;\omega)\,\frac{\omega^2}{c^2}\,\delta_{\mu\nu}-\frac{\partial^2}{\partial x_{\mu}\partial x_{\nu}}
+\delta_{\mu\nu}\,\nabla^2_{\bf r}\right]E^{\rm inc}_{\nu}({\mbox{\boldmath$r$}};\omega)=0\ .
\label{e2.5}
\end{equation}
The source term ${\cal P}_{\nu}^{\rm loc}({\bf r}^\prime;\omega)$ in Eq.\,(\ref{e2.4}) 
usually depends linearly on the total electric field (assuming a weak field) and can be related 
to the optical response function\,\cite{ref1,new} of an electronic system.
\medskip

Specifically, for a translationally invariant hybrid semi-infinite system, the Green's 
function can be expressed in terms of its two-dimensional (2D) Fourier transform for the $(x_1,x_2)$-plane

\begin{equation}
{\cal G}_{\mu\nu}({\mbox{\boldmath$r$}},{\mbox{\boldmath$r$}}^\prime;\omega)=\int 
\frac{d^2{\mbox{\boldmath$q$}}_{\|}}{(2\pi)^2}\,e^{i{\bf q}_{\|}\cdot({\bf r}_{\|}-{\bf r}^\prime_{\|})}\,g_{\mu\nu}({\mbox{\boldmath$q$}}_{\|},\omega\vert x_3,x_3^\prime)\ ,
\label{e2.6}
\end{equation}
where we have introduced a 2D wave vector $\mbox{\boldmath$q$}_{\|}=(q_1,q_2)$.
Substituting Eq.\,(\ref{e2.6}) into Eq.\,(\ref{e2.3}), we obtain a set of coupled 
differential equations

\begin{equation}
{\footnotesize
\left[\begin{array}{ccc}
\displaystyle{\epsilon_b\,\frac{\omega^2}{c^2}-q_2^2+\frac{d^2}{dx_3^2}} 
& q_1q_2 & \displaystyle{-iq_1\frac{d}{dx_3}}\\
q_1q_2 & \displaystyle{\epsilon_b\,\frac{\omega^2}{c^2}-q_1^2+\frac{d^2}{dx_3^2}} 
& \displaystyle{-iq_2\frac{d}{dx_3}}\\
\displaystyle{-iq_1\frac{d}{dx_3}} & \displaystyle{-iq_2\frac{d}{dx_3}} 
& \displaystyle{\epsilon_b\,\frac{\omega^2}{c^2}-q_{\|}^2}
\end{array}\right]\,\left[\begin{array}{ccc}
g_{11} & g_{12} & g_{13}\\
g_{21} & g_{22} & g_{23}\\
g_{31} & g_{32} & g_{33}
\end{array}\right]
=\delta(x_3-x_3^\prime)\,\left[\begin{array}{ccc}
1 & 0 & 0\\
0 & 1 & 0\\
0 & 0 & 1
\end{array}\right]}\ .
\label{e2.7}
\end{equation}
After introducing a rotational transformation\,\cite{add11} in $\mbox{\boldmath$q$}_{\|}$-space

\begin{equation}
f_{\mu\nu}(q_{\|},\omega\vert x_3,x_3^\prime)
=\sum\limits_{\mu^\prime,\nu^\prime}\,{\cal S}_{\mu\mu^\prime}({\mbox{\boldmath$q$}}_{\|})\,{\cal S}_{\nu\nu^\prime}({\mbox{\boldmath$q$}}_{\|})\,g_{\mu^\prime\nu^\prime}({\mbox{\boldmath$q$}}_{\|},\omega\vert x_3,x_3^\prime)\ ,
\label{e2.8}
\end{equation}
where the rotational matrix ${\cal S}({\mbox{\boldmath$q$}}_{\|})$ is

\begin{equation}
{\cal S}({\mbox{\boldmath$k$}}_{\|})=\frac{1}{q_{\|}}\,\left[\begin{array}{ccc}
q_1 & q_2 & 0\\
-q_2 & q_1 & 0\\
0 & 0 & q_{\|}
\end{array}\right]\ ,
\label{e2.9}
\end{equation}
we get an equivalent but simple expression for Eq.\,(\ref{e2.7}) as

\begin{equation}
\left[\begin{array}{ccc}
\displaystyle{\epsilon_b\,\frac{\omega^2}{c^2}+\frac{d^2}{dx_3^2}} & 0 
& \displaystyle{-iq_{\|}\frac{d}{dx_3}}\\
0 & \displaystyle{\epsilon_b\,\frac{\omega^2}{c^2}-q_{\|}^2+\frac{d^2}{dx_3^2}} & 0\\
\displaystyle{-iq_{\|}\frac{d}{dx_3}} & 0 & \displaystyle{\epsilon_b\,\frac{\omega^2}{c^2}-q_{\|}^2}
\end{array}\right]\,\left[\begin{array}{ccc}
f_{11} & f_{12} & f_{13}\\
f_{21} & f_{22} & f_{23}\\
f_{31} & f_{32} & f_{33}
\end{array}\right]
=\delta(x_3-x_3^\prime)\,\left[\begin{array}{ccc}
1 & 0 & 0\\
0 & 1 & 0\\
0 & 0 & 1
\end{array}\right]\ .
\label{e2.10}
\end{equation}
In order to acquire the solution for Eq.\,(\ref{e2.10}), we have to apply both the 
finite-value boundary condition at $x_3^\prime=\pm\infty$ as well as the continuity 
boundary condition at the $x_3=0$ interface.
This leads to the following
five nonzero $f_{\mu\nu}(q_{\|},\omega\vert x_3,x_3^\prime)$ functions\,\cite{add11,add15} 
for Eq.\,(\ref{e2.8}), i.e., 

\begin{eqnarray}
&&f_{22}(q_{\|},\omega\vert x_3,x_3^\prime)
\nonumber\\
&=&\left\{\begin{array}{llll}
\displaystyle{-\left(\frac{i}{2p}\right)\frac{2p}{p_{d}+p}\ 
e^{ip_{d}x_3-ipx_3^\prime}}\ , & x_3>0\ ,\,x_3^\prime<0\\
\\
\displaystyle{-\left(\frac{i}{2p}\right)\left[e^{ip|x_3-x_3^\prime|}-
\frac{p_{d}-p}{p_{d}+p}\ e^{-ip(x_3+x_3^\prime)}\right]}\ , & x_3<0\ ,\,x_3^\prime<0\\
\\
\displaystyle{-\left(\frac{i}{2p_{d}}\right)\left[e^{ip_{d}|x_3-x_3^\prime|}+
\frac{p_{d}-p}{p_{d}+p}\ e^{ip_{d}(x_3+x_3^\prime)}\right]}\ , & x_3>0\ ,\,x_3^\prime>0\\
\\
\displaystyle{-\left(\frac{i}{2p_{d}}\right)\frac{2p_{d}}{p_{d}+p}\ e^{-ip(x_3-x_3^\prime)}}\ , 
& x_3<0\ ,\,x_3^\prime>0
\end{array}\right.\ ,\ \
\label{e2.11}
\end{eqnarray}

\begin{eqnarray}
&&f_{13}(q_{\|},\omega\vert x_3,x_3^\prime)
\nonumber\\
&=&\left\{\begin{array}{llll}
\displaystyle{\frac{iq_{\|}c^2}{2\epsilon_{M}(\omega)\omega^2}
\left[\frac{2\epsilon_{M}(\omega)p_{d}}{\epsilon_{M}(\omega)p_{d}+\epsilon_{d}\,p}\right]
\,e^{ip_{d}x_3-ipx_3^\prime}}\ , & x_3>0\ ,\,x_3^\prime<0\\
\\
\displaystyle{\frac{iq_{\|}c^2}{2\epsilon_{M}(\omega)\omega^2}\left[e^{ip|x_3-x_3^\prime|}\,{\rm sgn}(x_3-x_3^\prime)+\frac{\epsilon_{M}(\omega)p_{d}-\epsilon_{d}\,p}{\epsilon_{M}(\omega)p_{d}+\epsilon_{d}\,p}\ e^{-ip(x_3+x_3^\prime)}\right]}\ , & x_3<0\ ,\,x_3^\prime<0\\
\\
\displaystyle{\frac{iq_{\|}c^2}{2\epsilon_{d}\,\omega^2}\left[e^{ip_{d}|x_3-x_3^\prime|}\,{\rm sgn}(x_3-x_3^\prime)+\frac{\epsilon_{M}(\omega)p_{d}-\epsilon_{d}\,p}{\epsilon_{M}(\omega)p_{d}+\epsilon_{d}\,p}\
e^{ip_{d}(x_3+x_3^\prime)}\right]}\ , & x_3>0\ ,\,x_3^\prime>0\\
\\
\displaystyle{-\frac{iq_{\|}c^2}{2\epsilon_{d}\,\omega^2}\left[\frac{2\epsilon_{d}\,p}{\epsilon_{M}(\omega)p_{d}+\epsilon_{d}\,p}\right]\,e^{-ipx_3+ip_{d}x_3^\prime}}\ , & x_3<0\ ,\,x_3^\prime>0
\end{array}\right.
\label{e2.12}
\end{eqnarray}

\begin{eqnarray}
&&f_{33}(q_{\|},\omega\vert x_3,x_3^\prime)
\nonumber\\
&=&\left\{\begin{array}{llll}
\displaystyle{-\frac{ik^2_{\|}c^2}{\omega^2}\left[
\frac{1}{\epsilon_{M}(\omega)p_{d}+\epsilon_{d}\,p}\right]\,e^{ip_{d}x_3-ipx_3^\prime}}\ , 
& x_3>0\ ,\,x_3^\prime<0\\
\\
\displaystyle{\frac{c^2}{\epsilon_{M}(\omega)\omega^2}\,\delta(x_3-x_3^\prime)-
\frac{ik^2_{\|}c^2}{2p\,\epsilon_{M}(\omega)\omega^2}}\\
\displaystyle{\times\left[e^{ip|x_3-x_3^\prime|}-\frac{\epsilon_{M}(\omega)p_{d}-\epsilon_{d}\,p}
{\epsilon_{M}(\omega)p_{d}+\epsilon_{d}\,p}\ e^{-ip(x_3+x_3^\prime)}\right]}\ , & x_3<0\ ,\,x_3^\prime<0\\
\\
\displaystyle{\frac{c^2}{\epsilon_{d}\,\omega^2}\,\delta(x_3-x_3^\prime)-\frac{iq_{\|}^2c^2}{2p_{d}\epsilon_{d}\,\omega^2}}\\
\displaystyle{\times\left[e^{ip_{d}|x_3-x_3^\prime|}+\frac{\epsilon_{M}(\omega)p_{d}-\epsilon_{d}\,p}{\epsilon_{M}(\omega)p_{d}+\epsilon_{d}\,p}\ e^{ip_{d}(x_3+x_3^\prime)}\right]}\ , & x_3>0\ ,\,x_3^\prime>0\\
\\
\displaystyle{-\frac{ik^2_{\|}c^2}{\omega^2}\left[\frac{1}{\epsilon_{M}(\omega)p_{d}+\epsilon_{d}\,p}\right]\,e^{-ipx_3+ip_{d}x_3^\prime}}\ , & x_3<0\ ,\,x_3^\prime>0
\end{array}\right.
\label{e2.13}
\end{eqnarray}

\begin{eqnarray}
&&f_{11}(q_{\|},\omega\vert x_3,x_3^\prime)
\nonumber\\
&=&\left\{\begin{array}{llll}
\displaystyle{-\frac{ip_{d}p\,c^2}{\omega^2}\left[\frac{1}{\epsilon_{M}(\omega)p_{d}+\epsilon_{d}\,p}\right]\,e^{ip_{d}x_3-ipx_3^\prime}}\ , & x_3>0\ ,\,x_3^\prime<0\\
\\
\displaystyle{-\frac{ip\,c^2}{2\epsilon_{M}(\omega)\omega^2}\left[e^{ip|x_3-x_3^\prime|}+\frac{\epsilon_{M}(\omega)p_{d}-\epsilon_{d}\,p}{\epsilon_{M}(\omega)p_{d}+\epsilon_{d}\,p}\ e^{-ip(x_3+x_3^\prime)}\right]}\ , & x_3<0\ ,\,x_3^\prime<0\\
\\
\displaystyle{-\frac{ip_{d}c^2}{2\epsilon_{d}\,\omega^2}\left[e^{ip_{d}|x_3-x_3^\prime|}-\frac{\epsilon_{M}(\omega)p_{d}-\epsilon_{d}\,p}{\epsilon_{M}(\omega)p_{d}+\epsilon_{d}\,p}\ e^{ip_{d}(x_3+x_3^\prime)}\right]}\ , & x_3>0\ ,\,x_3^\prime>0\\
\\
\displaystyle{-\frac{ip_{d}c^2}{2\epsilon_{d}\,\omega^2}\left[\frac{2\epsilon_{d}p}{\epsilon_{M}(\omega)p_{d}+\epsilon_{d}\,p}\right]\,e^{-ipx_3+ip_{d}x_3^\prime}}\ , & x_3<0\ ,\,x_3^\prime>0
\end{array}\right.
\label{e2.14}
\end{eqnarray}

\begin{eqnarray}
&&f_{31}(q_{\|},\omega\vert x_3,x_3^\prime)
\nonumber\\
&=&\left\{\begin{array}{llll}
\displaystyle{\frac{iq_{\|}c^2}{\omega^2}\left[\frac{p}{\epsilon_{M}(\omega)p_{d}+\epsilon_{d}\,p}\right]\,e^{ip_{d}x_3-ipx_3^\prime}}\ , & x_3>0\ ,\,x_3^\prime<0\\
\\
\displaystyle{\frac{iq_{\|}c^2}{2\epsilon_{M}(\omega)\omega^2}\left[e^{ip|x_3-x_3^\prime|}\,{\rm sgn}(x_3-x_3^\prime)-\frac{\epsilon_{M}(\omega)p_{d}-\epsilon_{d}\,p}{\epsilon_{M}(\omega)p_{d}+\epsilon_{d}\,p}\ e^{-ip(x_3+x_3^\prime)}\right]}\ , & x_3<0\ ,\,x_3^\prime<0\\
\\
\displaystyle{\frac{iq_{\|}c^2}{2\epsilon_{d}\,\omega^2}\left[e^{ip_{d}|x_3-x_3^\prime|}\,{\rm sgn}(x_3-x_3^\prime)-\frac{\epsilon_{M}(\omega)p_{d}-\epsilon_{d}\,p}{\epsilon_{M}(\omega)p_{d}+\epsilon_{d}\,p}\
e^{ip_{d}(x_3+x_3^\prime)}\right]}\ , & x_3>0\ ,\,x_3^\prime>0\\
\\
\displaystyle{-\frac{iq_{\|}c^2}{\omega^2}\left[\frac{p_{d}}{\epsilon_{M}(\omega)p_{d}+\epsilon_{d\,}p}\right]\,e^{-ipx_3+ip_{d}x_3^\prime}}\ , & x_3<0\ ,\,x_3^\prime>0
\end{array}\right.
\label{e2.15}
\end{eqnarray}
where ${\rm sgn}(x)$ is the sign function,

\begin{equation}
p_{d}(q_{\|},\omega)=\sqrt{\epsilon_{d}\,\frac{\omega^2}{c^2}-q_{\|}^2}\ ,
\label{e2.16}
\end{equation}

\begin{equation}
p(q_{\|},\omega)=\sqrt{\epsilon_{M}(\omega)\,\frac{\omega^2}{c^2}-q_{\|}^2}\ ,
\label{e2.17}
\end{equation}
${\rm Im}[p_{d}(q_{\|},\omega)]\geq 0$, and ${\rm Im}[p(q_{\|},\omega)]\geq 0$.
From these five nonzero $f_{\mu\nu}(q_{\|},\omega\vert x_3,x_3^\prime)$ functions, we arrive at

\begin{equation}
g_{\mu\nu}({\mbox{\boldmath$q$}}_{\|},\omega\vert x_3,x_3^\prime)
=\sum\limits_{\mu^\prime,\nu^\prime}\,f_{\mu^\prime\nu^\prime}(q_{\|},\omega\vert x_3,x_3^\prime)\,{\cal S}_{\mu^\prime\mu}({\mbox{\boldmath$q$}}_{\|})\,{\cal S}_{\nu^\prime\nu}({\mbox{\boldmath$q$}}_{\|})\ ,
\label{e2.18}
\end{equation}
which can be substituted into Eq.\,(\ref{e2.6}) to obtain the Green's function 
${\cal G}_{\mu\nu}({\mbox{\boldmath$r$}},{\mbox{\boldmath$r$}}^\prime;\omega)$ in 
position space. However, in  our model system depicted in Fig.\,\ref{f1},  we only 
consider the case when $x_3,\,x'_3> 0$.

\section{Optical Response Function for Graphene}
\label{sec-3}

For an embedded 2D graphene sheet, the optical response function 
for Dirac electrons is found to be\,\cite{ref3}

\begin{equation}
\chi^{(0)}_{\rm s}(p_\|,\omega)=\left(\frac{e^2}{\epsilon_0\,p_\|^2}\right)
\Pi^{(0)}_{\rm s}(p_\|,\omega)\ ,
\label{b1}
\end{equation}
where ${\mbox{\boldmath$p$}}_\|=(p_1,p_2)$ stands for the in-plane electron 
wave vector, and $\Pi^{(0)}_{\rm s}(p_\|,\,\omega)$ represents the density-density 
response function for Dirac electrons within the graphene sheet which is given by\,\cite{add16}

\[
\Pi^{(0)}_{\rm s}(p_\|,\omega)=\frac{4}{{\cal A}}\sum_{n_1,n_2=\pm 1}
\sum_{{\bf k}_\|}\left|<n_1,{\mbox{\boldmath$k$}}_\|\vert e^{-i{\bf p}_\|
\cdot{\bf r}_\|}\vert n_2,{\mbox{\boldmath$k$}}_\|+{\mbox{\boldmath$p$}}_\|>\right|^2
\]
\begin{equation}
\times\frac{f_0(\varepsilon_{n_1,k_\|})-f_0(\varepsilon_{n_2,k_\|+p_\|})}{\varepsilon_{n_2,k_\|+p_\|}-\varepsilon_{n_1,k_\|}-\hbar(\omega+i0^+)}\ .
\label{e3.20}
\end{equation}
In Eq.\,(\ref{e3.20}), ${\cal A}$ is the normalization area for graphene,  ,
$\varepsilon_{\pm,k_\|}=\pm\sqrt{\hbar^2v_{F}^2k_\|^2+\varepsilon^2_{G}/4}$ are the 
kinetic energies for the upper ($+$, electrons) and lower ($-$, holes) Dirac cones, 
$v_{F}$ is the Fermi velocity of graphene electrons, $\varepsilon_{G}$ is the induced 
energy gap of the graphene sheet, and $f_0(x)$ represents  the Fermi-Dirac distribution 
function for thermal-equilibrium electrons. At very low temperatures, we have 
$f_0(\varepsilon_{n,k_\|})\approx\Theta(E_{F}-\varepsilon_{n,k_\|})$, where 
$E_{F}$ is the Fermi energy of doped electrons and $\Theta(x)$ is the unit step function.
\medskip

After a lengthy calculation, from Eq.\,(\ref{e3.20}) we obtain an analytic 
expression for a gapped graphene sheet at $T\approx 0$\,K as follows\,\cite{add16}:

\begin{eqnarray}
&&\Pi^{(0)}_{\rm s}(p_\|,\omega)=\frac{2E_{F}}{\pi\hbar^2v_F^2}-
\frac{p_\|^2}{4\pi\hbar\sqrt{|v_F^2p_\|^2-\omega^2|}}
\nonumber\\
&\times& \left\{i\left[G_>(x_{1,-})-G_>(x_{1,+})\right]{\cal Q}_{1_<}(x_{2,-})+
\left[G_<(x_{1,-})+iG_>(x_{1,+})\right]{\cal Q}_{2_<}(x_{2,-},\,x_{2,+})\right.
\nonumber\\
&+& \left[G_<(x_{1,+})+G_<(x_{1,-})\right]{\cal Q}_{3_<}(x_{2,-})+
\left[G_<(x_{1,-})-G_<(x_{1,+})\right]{\cal Q}_{4_<}(x_{2,+})
\nonumber\\
&+& \left[G_>(x_{1,+})-G_>(x_{1,-})\right]{\cal Q}_{1_>}(x_{2,-},\,x_3)
+\left[G_>(x_{1,+})+iG_<(x_{1,-})\right]{\cal Q}_{2_>}(x_{2,-},\,x_{2,+})
\nonumber\\
&+& \left[G_>(x_{1,+})-G_>(-x_{1.-})-i\pi[2-x_0^2]\right]{\cal Q}_{3_>}(x_{2,+})
\nonumber\\
&+& \left[G_>(-x_{1,-})+G_>(x_{1,+})-i\pi[2-x_0^2]\right]{\cal Q}_{4_>}(x_{2,-},\,x_3)
\nonumber\\
&+&\left.\left[G_0(x_{1,+})-G_0(x_{1,-})\right]{\cal Q}_{5_>}(x_3)\right\}\ ,
\label{b2}
\end{eqnarray}
where $E_{F}=\sqrt{(\hbar v_{F}k_{F})^2+(\varepsilon_{G}/2)^2}-\varepsilon_{G}/2$ with 
respect to the zero-energy point at $k_\|=0$, and $k_{F}=\sqrt{(E_{F}+
\varepsilon_{G}/2)^2-(\varepsilon_{G}/2)^2}/\hbar v_{F}$ is the Fermi wave number.
\medskip

In Eq.\,(\ref{b2}), we have introduced three self-defined functions which are given by

\begin{eqnarray}
&&G_<(x)=x\sqrt{x_0^2-x^2}-\left(2-x_0^2\right)\,\cos^{-1}\left(\frac{x}{x_0}\right)\ , \label{b3}\\
&&G_>(x)=x\sqrt{x^2-x_0^2}-\left(2-x_0^2\right)\,\cosh^{-1}\left(\frac{x}{x_0}\right)\ , \label{b4}\\
&&G_0(x)=x\sqrt{x^2-x_0^2}-\left(2-x_0^2\right)\,\sinh^{-1}\left(\frac{x}{\sqrt{-x_0^2}}\right)\ .
\label{b5}
\end{eqnarray}
Moreover, nine region functions employed in Eq.\,(\ref{b2}) are defined by

\begin{eqnarray}
&&{\cal Q}_{1_<}(x_{2,-})=\Theta(E_{F}-x_{2,-}-\hbar\omega)\ ,
\nonumber\\
&&{\cal Q}_{2_<}(x_{2,-},\,x_{2,+})=\Theta(-\hbar\omega-E_{F}+x_{2,-})\,\Theta(\hbar\omega+E_{F}-x_{2,-})\,\Theta(E_{F}+x_{2,+}-\hbar\omega)\ ,
\nonumber\\
&&{\cal Q}_{3_<}(x_{2,-})=\Theta(-E_{F}+x_{2,-}-\hbar\omega)\ ,
\nonumber\\
&&{\cal Q}_{4_<}(x_{2,+})=\Theta(\hbar\omega+E_{F}-x_{2,+})\,\Theta(\hbar v_{F}p_\|-\hbar\omega)\ ,
\nonumber\\
&&{\cal Q}_{1_>}(x_{2,-},\,x_3)=\Theta(2k_{F}-p_\|)\,\Theta(\hbar\omega-x_3)\,\Theta(E_{F}+x_{2,-}-\hbar\omega)\ ,
\nonumber\\
&&{\cal Q}_{2_>}(x_{2,-},\,x_{2,+})=\Theta(\hbar\omega-E_{F}-x_{2,-})\,\Theta(E_{F}+x_{2,+}-\hbar\omega)\ ,
\nonumber\\
&&{\cal Q}_{3_>}(x_{2,+})=\Theta(\hbar\omega-E_{F}-x_{2,+})\ ,
\nonumber\\
&&{\cal Q}_{4_>}(x_{2,-},\,x_3)=\Theta(p_\|-2k_{F})\,\Theta(\hbar\omega-x_3)\,\Theta(E_{F}+x_{2,-}-\hbar\omega)\ ,
\nonumber\\
&&{\cal Q}_{5_>}(x_3)=\Theta(\hbar\omega-\hbar v_{F}p_\|)\,\Theta(x_3-\hbar\omega)\ .
\end{eqnarray}
Finally, we have defined six
variables $x_0,\,x_{1,\pm},\,x_{2,\pm}$ and $x_3$ in region functions through

\begin{eqnarray}
&&x_0=\sqrt{1+\frac{\varepsilon_{G}^2}{\hbar^2v_{F}^2p_\|^2-\hbar^2\omega^2}}\ ,
\nonumber\\
&&x_{1,\pm}=\frac{2E_{F}\pm\hbar\omega}{\hbar v_{F}p_\|}\ ,
\nonumber\\
&&x_{2,\pm}=\sqrt{\hbar^2v_{F}^2(p_\|\pm k_{F})^2+\varepsilon_{G}^2/4}\ ,
\nonumber\\
&&x_3=\sqrt{\hbar^2v_{F}^2p_\|^2+\varepsilon_{G}^2}\ .
\end{eqnarray}
\medskip

For the gapless graphene sheet with $\varepsilon_{G}=0$, Eq.\,(\ref{b2}) reduces to\,\cite{zero}

\[
\Pi^{(0)}_{\rm s}(p_\|,\omega)=i\pi\,\frac{F(p_\|,\omega)}{\hbar^2v_{F}^2}+
\frac{2E_{F}}{\pi\hbar^2v_{F}^2}
-\frac{F(p_\|,\omega)}{\hbar^2v_{F}^2}\left\{G\left(\frac{\hbar
\omega+2E_{F}}{\hbar v_{F}p_\|}\right)-\Theta\left(\frac{2E_{F}-
\hbar\omega}{\hbar v_{F}p_\|}-1\right)\right.
\]
\begin{equation}
\left.\times\left[G\left(\frac{2E_{F}-\hbar\omega}{\hbar v_{F}p_\|}\right)
-i\pi\right]-\Theta\left(\frac{\hbar\omega-2E_{F}}{\hbar v_{F}p_\|}+1\right)
G\left(\frac{\hbar\omega-2E_{F}}{\hbar v_{F}p_\|}\right)\right\}\ ,
\label{b19}
\end{equation}
where another two self-defined functions are

\begin{equation}
F(p_\|,\omega)=\frac{1}{4\pi}\,\frac{\hbar v_{F}^2p_\|^2}{\sqrt{\omega^2-v_{F}^2p_\|^2}}\ ,
\label{b20}
\end{equation}

\begin{equation}
G(z)=z\sqrt{z^2-1}-\ln\left(z+\sqrt{z^2-1}\right)\ .
\label{b21}
\end{equation}

\section{Hybridized Modes for Double-Layer Graphene}
\label{sec-4}

We would like to emphasize that our model system, illustrated in Fig.\,\ref{f1}, 
consists of a semi-infinite conducting  substrate along with a dielectric material  
with an embedded double-layer graphene above the conductor surface.
A surface-plasmon (SP) field is locally excited through a surface grating by normally-incident 
light outside the graphene region. This surface-propagating SP field further excites
Dirac electrons in the off-surface coupled pair of graphene sheets. As a result, the induced 
optical-polarization field from the excited Dirac electrons constitutes local resonant scattering 
sources to the Maxwell equation for the propagating SP field in the system.\,\cite{ref1,add12}
\medskip

Making use of the Green's function 
${\cal G}_{\mu\nu}({\mbox{\boldmath$r$}},{\mbox{\boldmath$r$}}^\prime;\omega)$ in 
Eq.\,(\ref{e2.6}), we have converted the Maxwell equation for the electric field 
$\mbox{\boldmath$E$}({\mbox{\boldmath$r$}};\omega)$ into a three-dimensional integral 
equation, as presented by Eq.\,(\ref{e2.4}) in which 
$\mbox{\boldmath$E$}^{\rm inc}({\mbox{\boldmath$r$}};\omega)$ represents the external 
SP near field in the region defined as  $x_3>0$, given explicitly by\,\cite{ref1,new}

\begin{equation}
\mbox{\boldmath$E$}^{\rm inc}({\mbox{\boldmath$r$}};\omega)=E_0\,
e^{i{\bf q}_0(\omega)\cdot{\bf D}_0}\,\frac{c}{\omega}
\left[i\hat{\mbox{\boldmath$q$}}_0\beta_3(q_0,\omega)-
\hat{\mbox{\boldmath$x$}}_3q_0(\omega)\right]\,e^{i{\bf q}_0(\omega)
\cdot{\bf r}_\|}\,e^{-\beta_3(q_0,\,\omega)x_3}\ .
\label{e2}
\end{equation}
In Eq.\,(\ref{e2}), $\hat{\mbox{\boldmath$q$}}_0$ and $\hat{\mbox{\boldmath$x$}}_3$ 
are unit vectors along the directions of the in-plane SP wave vector 
$\mbox{\boldmath$q$}_0(\omega)=q_0(\omega)(\cos\theta_0,\,\sin\theta_0)$
and $x_3$, $E_0$ is the field amplitude, $\omega$ is the field frequency, 
$\theta_0$ is the angle of the incident SP field with respect to the 
$x_1$ direction, $\mbox{\boldmath$D$}_0$ indicates the position vector
of the surface grating, and the introduced in-plane and out-of-plane wave 
numbers are given, respectively, by

\begin{equation}
q_0(\omega)=\frac{\omega}{c}\sqrt{\frac{\epsilon_{d}\,
\epsilon_{M}(\omega)}{\epsilon_{d}+\epsilon_{M}(\omega)}}\ ,
\label{e3}
\end{equation}

\begin{equation}
\beta_3(q_0,\omega)=\sqrt{q^2_0(\omega)-\frac{\omega^2}{c^2}}\ ,
\label{e4}
\end{equation}
where ${\rm Re}[q_0(\omega)]\geq 0$ and ${\rm Re}[\beta_3(q_0,\omega)]\geq 0$.
As $q_0\rightarrow\infty$, from Eq.\,(\ref{e3}) we know that $\epsilon_{d}+
\epsilon_{M}(\omega)=0$, which gives rise to the uncoupled SP energy\,\cite{sp} 
$\hbar\Omega_r=\hbar\Omega_p/\sqrt{\epsilon_s+\epsilon_d}$.
Therefore, Eq.\,(\ref{e2}) represents the SP-like near field in the limit of 
$q_0\rightarrow\infty$, while it becomes a light-like radiation field in the limit 
of $q_0\rightarrow 0$. The complex $\epsilon_{M}(\omega)$ in Eq.\,(\ref{e3}) implies 
an in-plane propagation loss for the SP field.
\medskip

For the two-dimensional graphene sheets, we can simply write down 
$\mbox{\boldmath${\cal P}$}^{\rm loc}({\mbox{\boldmath$r$}}^\prime;\omega)
=\sum\limits_{j=0,1}\,
\mbox{\boldmath${\cal P}$}^{\rm s}({\mbox{\boldmath$r$}}^\prime_\|;\omega
\vert z_j)\,\delta(x_3^\prime-z_j)$ with
$z_j$ labeling the positions of two graphene sheets in the $x_3$ direction. Therefore, 
from Eq.\,(\ref{e2.4}) we obtain

\begin{equation}
E_{\mu}({\mbox{\boldmath$r$}}_\|;\omega\vert x_3)=
E^{\rm inc}_{\mu}({\mbox{\boldmath$r$}}_\|;\omega\vert x_3)+\frac{\omega^2}{\epsilon_0c^2}
\sum\limits_{\nu=1}^{3}\,\sum_{j=0}^{1}\int d^2{\mbox{\boldmath$r$}}_\|^\prime\,
{\cal G}_{\mu\nu}({\mbox{\boldmath$r$}}_\|,{\mbox{\boldmath$r$}}_\|^\prime;\omega
\vert x_3,z_j)\,{\cal P}_{\nu}^{\rm s}({\mbox{\boldmath$r$}}_\|^\prime;\omega\vert z_j)\ ,
\label{e5}
\end{equation}
where ${\cal G}_{\mu\nu}({\mbox{\boldmath$r$}}_\|,{\mbox{\boldmath$r$}}_\|^\prime;\omega\vert x_3,z_j)$ represents ${\cal G}_{\mu\nu}({\mbox{\boldmath$r$}},{\mbox{\boldmath$r$}}^\prime;\omega)$ evaluated at ${\mbox{\boldmath$r$}}=({\mbox{\boldmath$r$}}_\|,x_3)$ $\&$
${\mbox{\boldmath$r$}}^\prime=({\mbox{\boldmath$r$}}^\prime_\|,z_j)$, and $\mbox{\boldmath$E$}({\mbox{\boldmath$r$}}_\|;\omega\vert x_3)$ is simply $\mbox{\boldmath$E$}({\mbox{\boldmath$r$}};\omega)$ at
${\mbox{\boldmath$r$}}=({\mbox{\boldmath$r$}}_\|,x_3)$.
\medskip

After performing a Fourier transformation on the Green's function, as given by Eq.\,(\ref{e2.6}), 
for the translationally invariant semi-infinite hybrid conductor system within 
the $(x_1,x_2)$-plane, we can rewrite Eq.\,(\ref{e5}) as

\[
E_{\mu}({\mbox{\boldmath$r$}}_\|;\omega\vert x_3)=E^{\rm inc}_{\mu}({\mbox{\boldmath$r$}}_\|;\omega\vert x_3)
\]
\begin{equation}
+\frac{\omega^2}{\epsilon_0c^2}\sum\limits_{\nu=1}^{3}\,\sum_{j=0}^{1}
\int \frac{d^2{\mbox{\boldmath$q$}}_{\|}}{(2\pi)^2}\,
e^{i{\bf q}_\|\cdot{\bf r}_\|}\,g_{\mu\nu}({\mbox{\boldmath$q$}}_\|,\omega\vert x_3,z_j)\,{\cal P}_{\nu}^{\rm s}({\mbox{\boldmath$q$}}_\|,\omega\vert z_j)\ ,
\label{e7}
\end{equation}
where we have introduced the Fourier transformed polarization field

\begin{equation}
\mbox{\boldmath${\cal P}$}^{\rm s}({\mbox{\boldmath$q$}}_\|,\omega\vert z_j)=\int d^2{\mbox{\boldmath$r$}}^\prime_\|\,e^{-i{\bf q}_\|\cdot{\bf r}^\prime_\|}\,\mbox{\boldmath${\cal P}$}^{\rm s}({\mbox{\boldmath$r$}}^\prime_\|;\omega\vert z_j)\ .
\label{e8}
\end{equation}
Using linear response theory\,\cite{new,ref2} for translationally-invariant monolayer
graphene sheets within the $(x_1,x_2)$-plane, we obtain

\begin{equation}
{\cal P}_{\nu}^{\rm s}({\mbox{\boldmath$q$}}_\|,\omega\vert z_j)=\epsilon_0\chi^{(0)}_j(q_\|,\omega)\left(1-\delta_{\nu 3}\right)\int d^2{\mbox{\boldmath$r$}}^\prime_\|\,e^{-i{\bf q}_\|\cdot{\bf r}^\prime_\|}\,
E_{\nu}({\mbox{\boldmath$r$}}^\prime_\|;\omega\vert z_j)\ ,
\label{e9}
\end{equation}
where the optical polarization of graphene is limited within each 
sheet, $\chi^{(0)}_j(q_\|,\omega)$ is the optical-response function for 
Dirac electrons within the $j$th graphene sheet and is given,
under the condition $q_\|>\omega/c$, by Eqs.\,(\ref{b1}) and (\ref{e3.20}).
\medskip

Setting $x_3=z_j$ in Eq.\,(\ref{e7}) and meanwhile using Eq.\,(\ref{e9}),
we arrive at the following two coupled self-consistent equations for the total electric 
field $\mbox{\boldmath$E$}({\mbox{\boldmath$q$}}_\|,\omega\vert z_j)$ on the graphene 
sheets, i.e.,

\[
E_{\mu}({\mbox{\boldmath$q$}}_\|,\omega\vert z_j)=E^{\rm inc}_{\mu}({\mbox{\boldmath$q$}}_\|,\omega\vert z_j)
\]
\begin{equation}
+\frac{\omega^2}{c^2}\sum\limits_{\nu=1}^{3}\,\sum_{j'=0}^{1}
g_{\mu\nu}({\mbox{\boldmath$q$}}_\|,\omega\vert z_j,z_{j'})\left(1-\delta_{\nu 3}\right)\chi^{(0)}_{j'}(q_\|,\omega)\,E_{\nu}({\mbox{\boldmath$q$}}_\|,\omega\vert z_{j'})\ ,
\label{e12}
\end{equation}
where $j=0,\,1$ and we have used a Fourier transform for the electric field 
$\mbox{\boldmath$E$}({\mbox{\boldmath$r$}}_\|;\omega\vert z_j)$ as

\begin{equation}
\mbox{\boldmath$E$}({\mbox{\boldmath$q$}}_\|,\omega\vert z_j)=\int d^2{\mbox{\boldmath$r$}}_\|\,e^{-i{\bf q}_\|\cdot{\bf r}_\|}\,\mbox{\boldmath$E$}({\mbox{\boldmath$r$}}_\|;\omega\vert z_j)\ .
\label{e13}
\end{equation}
Moreover, $\mbox{\boldmath$E$}^{\rm inc}({\mbox{\boldmath$q$}}_{\|},\omega\vert z_j)$ 
in Eq.\,(\ref{e12}) can be calculated directly from Eq.\,(\ref{e2}) as

\[
\mbox{\boldmath$E$}^{\rm inc}({\mbox{\boldmath$q$}}_{\|},\omega\vert z_j)=\delta({\mbox{\boldmath$q$}}_{\|}-{\mbox{\boldmath$q$}}_0)\,E_0\,e^{i{\bf q}_0\cdot{\bf D}_0}\,\frac{(2\pi)^2c}{\omega}
\left(i\hat{\mbox{\boldmath$q$}}_0\beta_3-\hat{\mbox{\boldmath$x$}}_3q_0\right)\,e^{-\beta_3z_j}
\]
\begin{equation}
\equiv(2\pi)^2\delta({\mbox{\boldmath$q$}}_{\|}-{\mbox{\boldmath$q$}}_0)\,\mbox{\boldmath$A$}({\mbox{\boldmath$q$}}_0,\omega\vert z_j)\,E_0\,e^{i{\bf q}_0\cdot{\bf D}_0}\ .
\label{e14}
\end{equation}
where $A_{\mu}({\mbox{\boldmath$q$}}_0,\omega\vert z_j)$ for $\mu=1,\,2,\,3$ represents 
the field enhancement factors.
\medskip

If we set $\mbox{\boldmath$E$}^{\rm inc}({\mbox{\boldmath$q$}}_{\|},\omega\vert z_j)=0$   
in Eq.\,(\ref{e12}), we are able to obtain the following dispersion equation 
for the self-sustained density oscillations
within two graphene sheets, and the resulting dispersion relation 
$\omega=\Omega_{\rm g-sp}({\mbox{\boldmath$q$}}_\|)$ for the hybrid 
graphene-surface plasmon modes is determined by the following secular equation\,\cite{ref2-1}

\begin{equation}
{\cal D}et\left[\delta_{\mu\nu}\,\delta_{jj'}-\frac{\omega^2}{c^2}\,
g_{\mu\nu}({\mbox{\boldmath$q$}}_\|,\omega\vert z_j,z_{j'})\left(1-\delta_{\nu 3}\right)\chi^{(0)}_{j'}(q_\|,\omega)\right]\equiv{\cal D}et\left[\tilde{\cal C}^{jj'}_{\mu\nu}({\mbox{\boldmath$q$}}_\|,\omega)\right]=0\ ,
\label{e15}
\end{equation}
where $\mu,\,\nu=1,\,2,\,3$, $j,j'=0,\,1$, and the $2\times 2$ block (or $6\times 6$) 
coefficient matrix $\tilde{\cal C}^{jj'}_{\mu\nu}({\mbox{\boldmath$q$}}_\|,\omega)$ is given by

\begin{equation}
{\footnotesize
\left[\begin{array}{cc}
\displaystyle{\delta_{\mu\nu}-\frac{\omega^2}{c^2}g_{\mu\nu}({\mbox{\boldmath$q$}}_\|,\omega\vert z_0,z_0)\left(1-\delta_{\nu 3}\right)\chi^{(0)}_{0}(q_\|,\omega)} & \displaystyle{-\frac{\omega^2}{c^2}
g_{\mu\nu}({\mbox{\boldmath$q$}}_\|,\omega\vert z_0,z_1)\left(1-\delta_{\nu 3}\right)\chi^{(0)}_{1}(q_\|,\omega)}\\
\displaystyle{-\frac{\omega^2}{c^2}g_{\mu\nu}({\mbox{\boldmath$q$}}_\|,\omega\vert z_1,z_0)\left(1-\delta_{\nu 3}\right)\chi^{(0)}_{0}(q_\|,\omega)} & \displaystyle{\delta_{\mu\nu}-\frac{\omega^2}{c^2}
g_{\mu\nu}({\mbox{\boldmath$q$}}_\|,\omega\vert z_1,z_1)\left(1-\delta_{\nu 3}\right)\chi^{(0)}_{1}(q_\|,\omega)}
\end{array}\right]}\ .
\label{e16}
\end{equation}
The $z_j$ position dependence in Eq.\,(\ref{e15}) reflects the distinctive near-field 
coupling\,\cite{ref1,new} between the surface plasmons and Dirac electrons in graphene.
Here, the factor $g_{\mu\nu}({\mbox{\boldmath$q$}}_\|,\omega\vert z_j,z_{j'})$ comes 
from the surface-plasmon response, while the other factor $\chi^{(0)}_{j}(q_\|,\omega)$ 
corresponds to the graphene optical response. Therefore, their product in 
Eq.\,(\ref{e15}) represents contributions to the hybrid graphene-surface plasmon modes.
The uncoupled surface-plasmon dispersion relation is included through 
$g_{\mu\nu}({\mbox{\boldmath$q$}}_\|,\omega\vert z_j,z_j)$ in an inexplicit way
for $q_\|\rightarrow\infty$. Moreover, the interlayer graphene coupling becomes 
insignificant as $q_\||z_1-z_0|\gg 1$. Graphically, the dispersion relation of such 
hybrid graphene-surface plasmon modes can be shown with the sign switching in the 
density plot for the real part of 
$1/{\cal D}et\left[\tilde{\cal C}^{jj'}_{\mu\nu}({\mbox{\boldmath$q$}}_\|,\omega)\right]$ within the ($\omega,q_\|$)-plane.
\medskip

By using the $6\times 6$ inverted coefficient matrix 
$\{\tilde{\cal C}^{jj'}_{\mu\nu}({\mbox{\boldmath$q$}}_\|,\omega)\}^{-1}$ 
calculated from Eq.\,(\ref{e16}), we can further compute the distribution of 
the total electric field (i.e., incident field plus the scattering field) by

\[
E_{\mu}({\mbox{\boldmath$q$}}_\|,\omega\vert x_3)=E^{\rm inc}_{\mu}({\mbox{\boldmath$q$}}_\|,\omega\vert x_3)+\frac{\omega^2}{c^2}\sum_{j=0}^{1}\,\chi^{(0)}_{j}(q_\|,\omega)
\]
\begin{equation}
\times\sum\limits_{\nu=1}^{3}\,\left\{g_{\mu\nu}({\mbox{\boldmath$q$}}_\|,\omega\vert x_3,z_j)\left(1-\delta_{\nu 3}\right)
\left[\sum_{\mu^\prime=1}^{3}\,\sum_{j'=0}^{1}\,\{\tilde{\cal C}^{jj'}_{\nu\mu^\prime}({\mbox{\boldmath$q$}}_\|,\omega)\}^{-1}\,E^{\rm inc}_{\mu^\prime}({\mbox{\boldmath$q$}}_\|,\omega\vert z_{j'})\right]\right\}\ ,
\label{e17}
\end{equation}
where the second term represents the contribution from the scattering field\,\cite{new}.
\medskip

It is clear from Eq.\,(\ref{e17}) that in the absence of a graphene sheet, 
i.e. $\chi^{(0)}_{j}(q_\|,\omega)=0$, the semi-infinite dielectric (with a 
relative dielectric constant $\epsilon_{d}$) is static, uniform and isotropic in 
the upper half space with an effective scattering matrix\,\cite{new} 
$\alpha_{\mu\nu}^{\rm eff}({\mbox{\boldmath$q$}}_\|,\omega\vert x_3,x_3^\prime)=0$. 
In the presence of the graphene sheets, on the other hand, the induced local polarization 
fields from the excited Dirac  electrons within the graphene sheets are introduced.
In addition, the retarded coulomb coupling between plasmon excitations in the graphene 
and in the semi-infinite conductor is also introduced into the system at the same time. 
Consequently, the effective scattering matrix 
$\alpha_{\mu\nu}^{\rm eff}({\mbox{\boldmath$q$}}_\|,\omega\vert x_3,x_3^\prime)$
becomes finite, dynamical, non-uniform and anisotropic in the space, and is 
calculated from Eq.\,(\ref{e17}) as 

\begin{eqnarray}
&&\alpha_{\mu\nu}^{\rm eff}({\mbox{\boldmath$q$}}_\|,\omega\vert x_3,x_3^\prime)\equiv
\frac{\partial}{\partial E^{\rm inc}_{\nu}({\mbox{\boldmath$q$}}_\|,\omega\vert x'_3)}
\left[E_{\mu}({\mbox{\boldmath$q$}}_\|,\omega\vert x_3)-E^{\rm inc}_{\mu}({\mbox{\boldmath$q$}}_\|,\omega\vert x_3)\right]
\nonumber
\\
&=&\frac{\omega^2}{c^2}\,\sum_{j'=0}^{1}\,\delta(x_3^\prime-z_{j'})\,\sum_{j=0}^{1}\,
\chi^{(0)}_{j}(q_\|,\omega)\sum\limits_{\nu^\prime=1}^{3}\,g_{\mu\nu^\prime}({\mbox{\boldmath$q$}}_\|,\omega\vert x_3,z_j)\left(1-\delta_{\nu^\prime 3}\right)
\{\tilde{\cal C}^{jj'}_{\nu^\prime\nu}({\mbox{\boldmath$q$}}_\|,\omega)\}^{-1}\ .\ \ \ \
\label{eff-1}
\end{eqnarray}
In Eq.\,(\ref{eff-1}), ${\mbox{\boldmath$q$}}_\|$ is a real vector, the single 
factor $\chi^{(0)}_{j}(q_\|,\omega)$
represents the contribution from the resonant excitation of Dirac electrons within 
the graphene sheet, while the combined factor
$g_{\mu\nu^\prime}({\mbox{\boldmath$q$}}_\|,\omega\vert x_3,z_j)\left(1-\delta_{\nu^\prime 3}\right)\{\tilde{\cal C}^{jj'}_{\nu^\prime\nu}({\mbox{\boldmath$q$}}_\|,\omega)\}^{-1}$
corresponds to the electromagnetic coupling between the semi-infinite conductor and the 
graphene sheet. Using Eq.\,(\ref{eff-1}), we can define a local effective scattering matrix 
through

\begin{eqnarray}
&&\alpha_{\mu\nu}^{\rm eff}({\mbox{\boldmath$q$}}_\|,\omega\vert x_3)=\int\limits_{0}^{\infty} dx_3^\prime\,\alpha_{\mu\nu}^{\rm eff}({\mbox{\boldmath$q$}}_\|,\omega\vert x_3,x_3^\prime)
\nonumber
\\
&=&\frac{\omega^2}{c^2}\,\sum_{j=0}^{1}\,
\chi^{(0)}_{j}(q_\|,\omega)\sum\limits_{\nu^\prime=1}^{3}\,g_{\mu\nu^\prime}({\mbox{\boldmath$q$}}_\|,\omega\vert x_3,z_j)\left(1-\delta_{\nu^\prime 3}\right)\,
\sum_{j'=0}^{1}\,\{\tilde{\cal C}^{jj'}_{\nu^\prime\nu}({\mbox{\boldmath$q$}}_\|,\omega)\}^{-1}\ .\ \ \ \
\label{eff-2}
\end{eqnarray}
which displays two peaks at $x_3=z_0,\,z_1$, and the broadening of the peak is determined 
by the exponential decay\,\cite{add11,add15} of the Green's function.
This implies that the dielectric constant in the region
between graphene sheets and the surface of the semi-infinite conductor will be 
modified significantly only if the graphene sheets stay very close to the surface 
of the semi-infinite conductor, i.e., the SP wavelength is required to be larger 
than the sheet separation from the surface.
\medskip

Using Eq.\,(\ref{e17}), we also get the total electric field in the real space, yielding

\[
E_{\mu}({\mbox{\boldmath$r$}}_\|,\omega\vert x_3)=E^{\rm inc}_{\mu}({\mbox{\boldmath$r$}}_\|,\omega\vert x_3)+\frac{\omega^2}{c^2}\int \frac{d^2{\mbox{\boldmath$q$}}_\|}{(2\pi)^2}\,
e^{i{\bf q}_\|\cdot{\bf r}_\|}\sum_{j=0}^{1}\,\chi^{(0)}_{j}(q_\|,\omega)
\]
\begin{equation}
\times\sum\limits_{\nu=1}^{3}\,\left\{g_{\mu\nu}({\mbox{\boldmath$q$}}_\|,\omega\vert x_3,z_j)\left(1-\delta_{\nu 3}\right)
\left[\sum_{\mu^\prime=1}^{3}\,\sum_{j'=0}^{1}\,\{\tilde{\cal C}^{jj'}_{\nu\mu^\prime}({\mbox{\boldmath$q$}}_\|,\omega)\}^{-1}\,E^{\rm inc}_{\mu^{\prime}}({\mbox{\boldmath$q$}}_\|,\omega\vert z_{j'})\right]\right\}\ ,
\label{e17r}
\end{equation}
where ${\mbox{\boldmath$E$}}^{\rm inc}({\mbox{\boldmath$r$}}_\|,\omega\vert x_3)$ 
has already been given by Eq.\,(\ref{e2}), and 
${\mbox{\boldmath$E$}}^s({\mbox{\boldmath$r$}}_\|,x_3\vert\omega)={\mbox{\boldmath$E$}}({\mbox{\boldmath$r$}}_\|,\omega\vert x_3)-{\mbox{\boldmath$E$}}^{\rm inc}({\mbox{\boldmath$r$}}_\|,\omega\vert x_3)$ 
stands for the spatial distribution of the scattering field.
\medskip

Furthermore, by employing the calculated electric field 
$\mbox{\boldmath$E$}({\mbox{\boldmath$q$}}_\|,\omega\vert z_j)$ 
on the graphene sheets from Eq.\,(\ref{e12}), the optical-absorption coefficient
$\beta_{\rm abs}(\omega)$ for the SP field by Dirac electrons is expressed 
as\,\cite{abs,lorentz,abs2,abs3}

\begin{equation}
\beta_{\rm abs}(\omega)=\frac{\omega\sqrt{\epsilon_{d}}}{n_r(\omega)\,c}\left[\frac{1}{\exp(\hbar\omega/k_{\rm
B}T)-1}+1\right]\,{\rm Im}\left[\alpha_{L}(\omega)\right]\ ,
\label{e18}
\end{equation}
where $\alpha_{L}(\omega)$ is the complex Lorentz function given by

\[
\alpha_{L}(\omega)=\left(\frac{2\pi e^2}{\epsilon_0\epsilon_{r}q_0^2}\right)\left(q_0^2-\epsilon_{d}\,\frac{\omega^2}{c^2}\right)^{1/2}\,\sum_{j=0}^{1}\,
\left[\Pi^{(0)}_{j}(q_0,\,\omega)+\{\Pi^{(0)}_{j}(q_0,\,-\omega)\}^\ast\right]
\]
\begin{equation}
\times\left|\sum_{\mu=1}^3\,\hat{e}_{\mu}\,\sum_{\nu=1}^3\,\sum_{j'=0}^{1}\,
\{\tilde{\cal C}^{jj'}_{\mu\nu}({\mbox{\boldmath$q$}}_0,\omega)\}^{-1}A_{\nu}({\mbox{\boldmath$q$}}_0,\omega\vert z_{j'})\right|\ ,
\label{e19}
\end{equation}
$\epsilon_r$ is the average dielectric constant of graphene embedded in the dielectric host.
In Eq.\,(\ref{e19}), ${\mbox{\boldmath$q$}}_0\equiv{\rm Re}[q_0(\omega)]\,\hat{\mbox{\boldmath$q$}}_0$ is a real vector,
$\hat{\mbox{\boldmath$e$}}^{\rm inc}=(\hat{e}_1,\hat{e}_2,\hat{e}_3)$ represents the unit polarization vector for the propagating SP field,
and the scaled refractive index function $n_r(\omega)$
in Eq.\,(\ref{e18}) is\,\cite{abs,lorentz,abs2,abs3}

\begin{equation}
n_r(\omega)=
\frac{1}{\sqrt{2}}\left(1+{\rm Re}\left[\alpha_{L}(\omega)\right]
+\sqrt{\left\{1+{\rm Re}\left[\alpha_{L}(\omega)\right]\right\}^2+\left\{{\rm Im}\left[\alpha_{L}
(\omega)\right]\right\}^2}\,\right)^{1/2}\ .
\label{e20}
\end{equation}
We now turn to presenting and discussing our numerical results in the next section. 

\section{Results and Discussions}
\label{sec-5}

In our numerical calculations,  we use the Fermi wave vector $k_{F}=\sqrt{\pi n_0}$ 
as the scale for wave numbers, $1/k_{F}$ for lengths, and $E_{F}=\hbar v_{F}k_{F}$ for 
energies. The direction of propagation of the SP field is chosen as
$\hat{\mbox{\boldmath$q$}}_0=(1,0,0)$ for convenience, and we also 
set $\epsilon_{s}=13.3$, $\epsilon_{d}=\epsilon_{r}=2.4$,
$v_{F}=1\times 10^8\,$cm/s, and $n_0=5\times 10^{11}\,$cm$^{-2}$ for the 
doping density in graphene. Moreover, the SP energy $\hbar\Omega_r$ and the half 
bandgap $\Delta=\varepsilon_G/2$ will be given directly in figure captions.
\medskip

\begin{figure}
\centering
\includegraphics[width=0.48\textwidth]{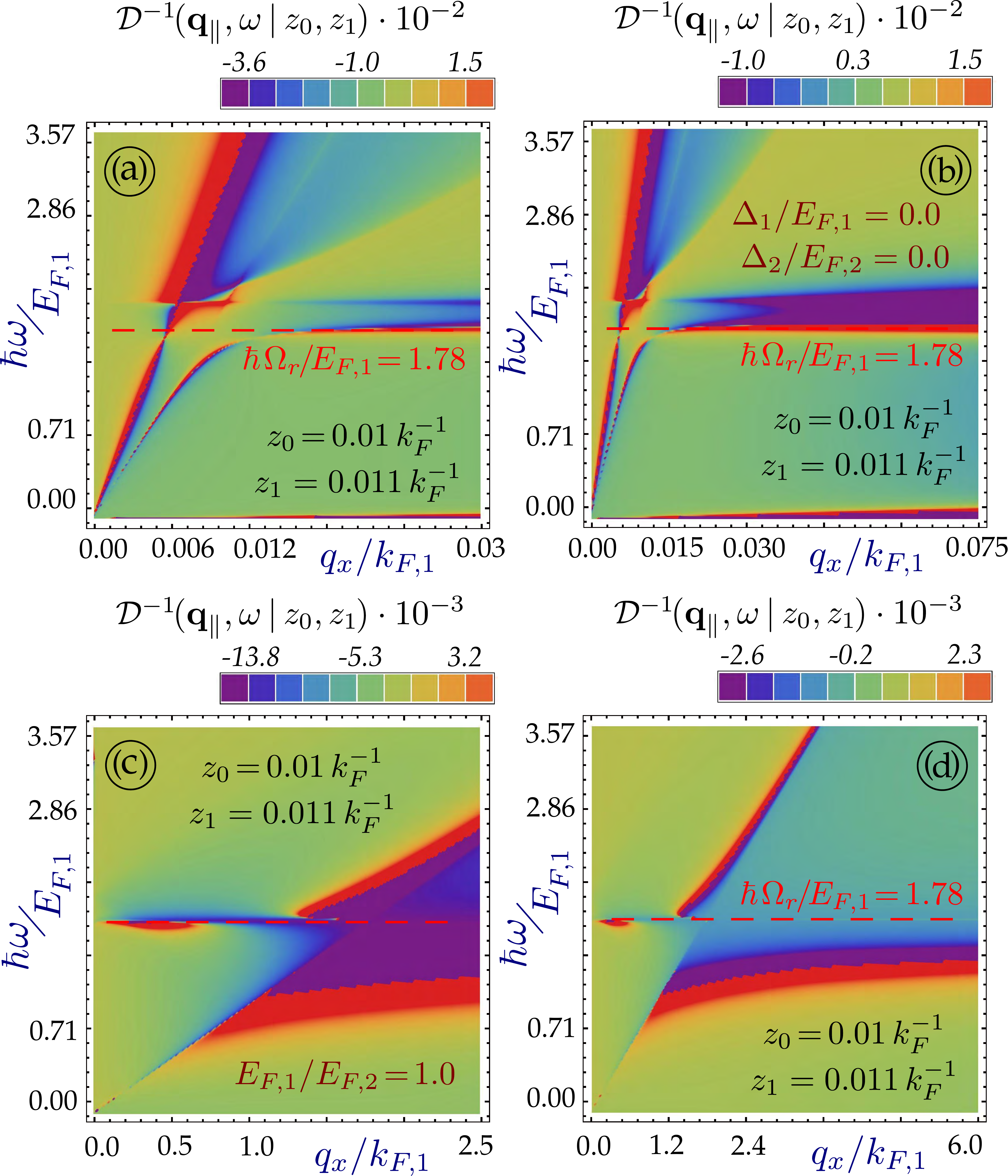}
\caption{(Color online)    Results for hybrid plasmon modes in different ranges of 
wavenumber. Density plots are presented for the real part of 
$\mathcal{D}^{-1}({\mbox{\boldmath$q$}}_\|, \omega \, 
\vert \, z_0, z_1)= 1/{\cal D}et\left[\tilde{\cal C}^{jj'}_{\mu\nu}(q_x,\omega)\right]$ 
using Eq.\,(\ref{e16})  with $q_x^{\rm max}/k_{F} = 0.03,\,0.075,\,2.5$ and 
$6.0$. The hybrid-plasmon dispersions initially appear 
as strong variations between positive (red) and negative (violet) 
peaks. The layer separations from the conductor surface are $z_0=0.01\,k_{F}^{-1}$ and
$z_1=0.011\,k_{F}^{-1}$. The surface-plasmon energy is equal to $\hbar\Omega_r =1.78\,E_{F,1}$.
Each graphene sheet is equally doped up to a Fermi energy $E_{F,1}=E_{F,2}=E_F$ and 
acquires a zero bandgap, i.e.,  $\Delta_1=\Delta_2=0$.}
\label{f3}
\end{figure}
For a retarded interaction between light and graphene electrons, both radiative and 
evanescent modes must be considered in the hybrid system. The radiative modes include 
photons and polaritons, while the evanescent (localized) modes appear as surface-plasmon 
polaritons (SPPs), graphene plasmons (G-Ps), and surface plasmons (SPs).
Figure\ \ref{f3} displays the real part of 
$\mathcal{D}^{-1}({\bf q}_\parallel, \omega \, \vert \, z_1, z_2)
= 1/{\cal D}et\left[\tilde{\cal C}^{jj'}_{\mu\nu}(q_x,\omega)\right]$
for four different ranges of $q_x$. As Fig.\,\ref{f3}($a$) shows, in addition to the SPP mode,
the hybridizations of both radiative photon and polariton modes with localized 
SPs (illustrated in Fig.\,\ref{f2}) appear in this very small $q_x$ range.
As the $q_x$ range slightly expands in Fig.\,\ref{f3}($b$), the SPP mode in 
Fig.\,\ref{f3}($a$) is fully developed, which is accompanied by
two degenerate acoustic-like G-P modes at very low energies. As the $q_x$ range 
further increases in Figs.\,\ref{f3}($c$) and \ref{f3}($d$), the G-P energy exceeds 
that of the SPP. Consequently, a single anticrossing of the G-P with SPP appears.
\medskip

\begin{figure}
\centering
\includegraphics[width=0.48\textwidth]{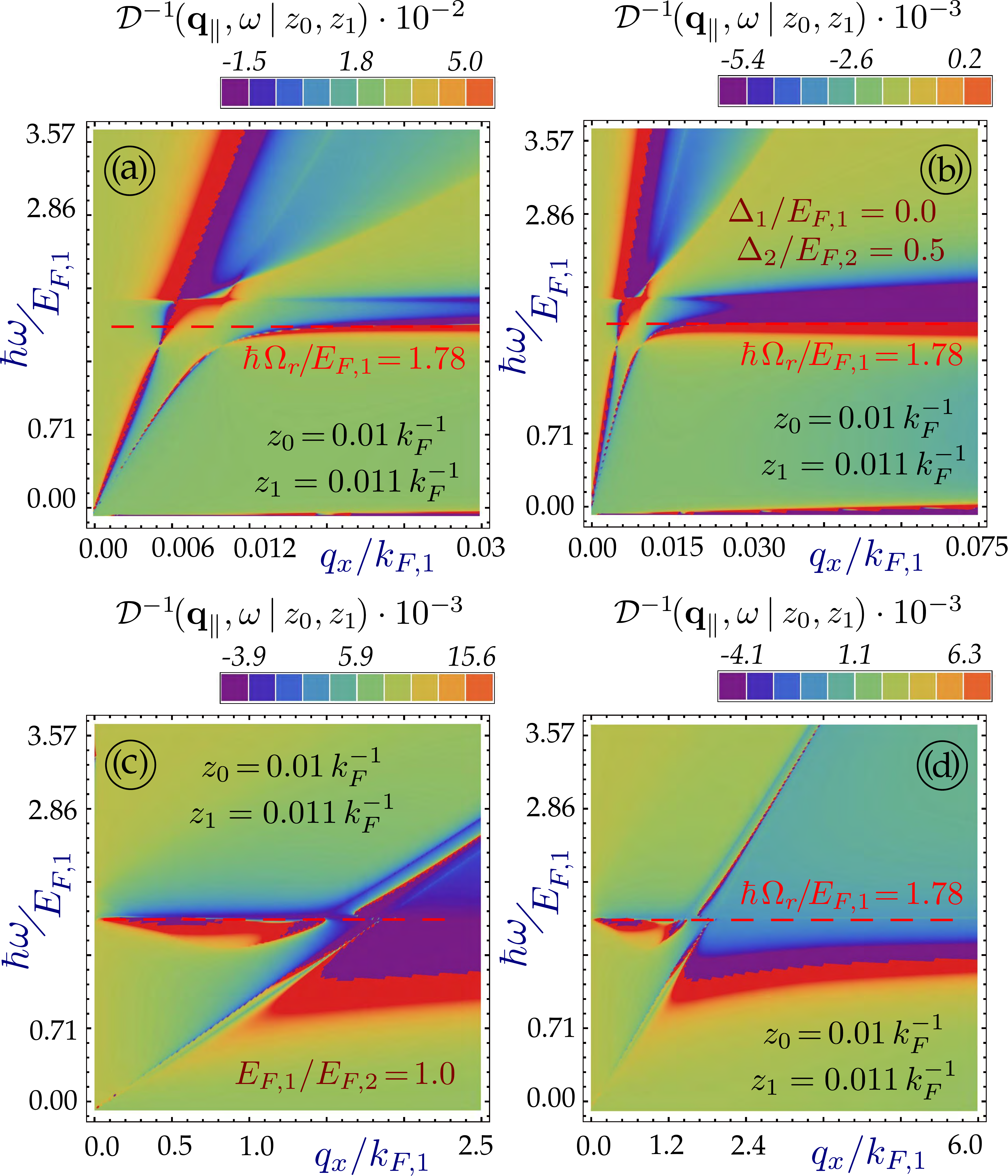}
\caption{(Color online)  Density plots for the real part of 
$\mathcal{D}^{-1}({\mbox{\boldmath$q$}}_\parallel, \omega \, \vert \, z_0, z_1)$
with $q_x^{\rm max}/k_{F} = 0.03,\,0.075,\,2.5$ and $6.0$.  We set $z_1=0.01\,k_{F}^{-1}$,
$z_2=0.011\,k_{F}^{-1}$ and $\hbar \Omega_r = 1.78\,E_{F}$.
Each graphene sheet has the same Fermi energy $E_{F,1}=E_{F,2}=E_F$ but has 
different bandgaps $\Delta_1=0$ and
$\Delta_2 = 0.5\,E_{F}$.}
\label{f4}
\end{figure}
In Fig.\,\ref{f4}, a finite bandgap  parameter $\Delta_2=0.5\,E_F$ is introduced to the top graphene 
layer, and two G-P modes become non-degenerate. By comparing with Fig.\,\ref{f3}, 
no changes in Figs.\,\ref{f4}($a$) and \ref{f4}($b$) are found for photon and polariton 
modes in smaller ranges of $q_x$. However,
as the $q_x$ range is increased in Figs.\,\ref{f4}($c$) and \ref{f4}($d$), 
the splitting of the two acoustic-like G-P modes and two optical-like SPP modes become 
visible in Fig.\,\ref{f4}($c$), where the lower (higher) energy G-P mode is associated 
with the top (bottom) layer. Moreover, the top-layer G-P mode after the second 
anticrossing in Fig.\,\ref{f4}($d$) is enhanced by reducing the Landau damping due 
to a finite bandgap.
\medskip

\begin{figure}
\centering
\includegraphics[width=0.48\textwidth]{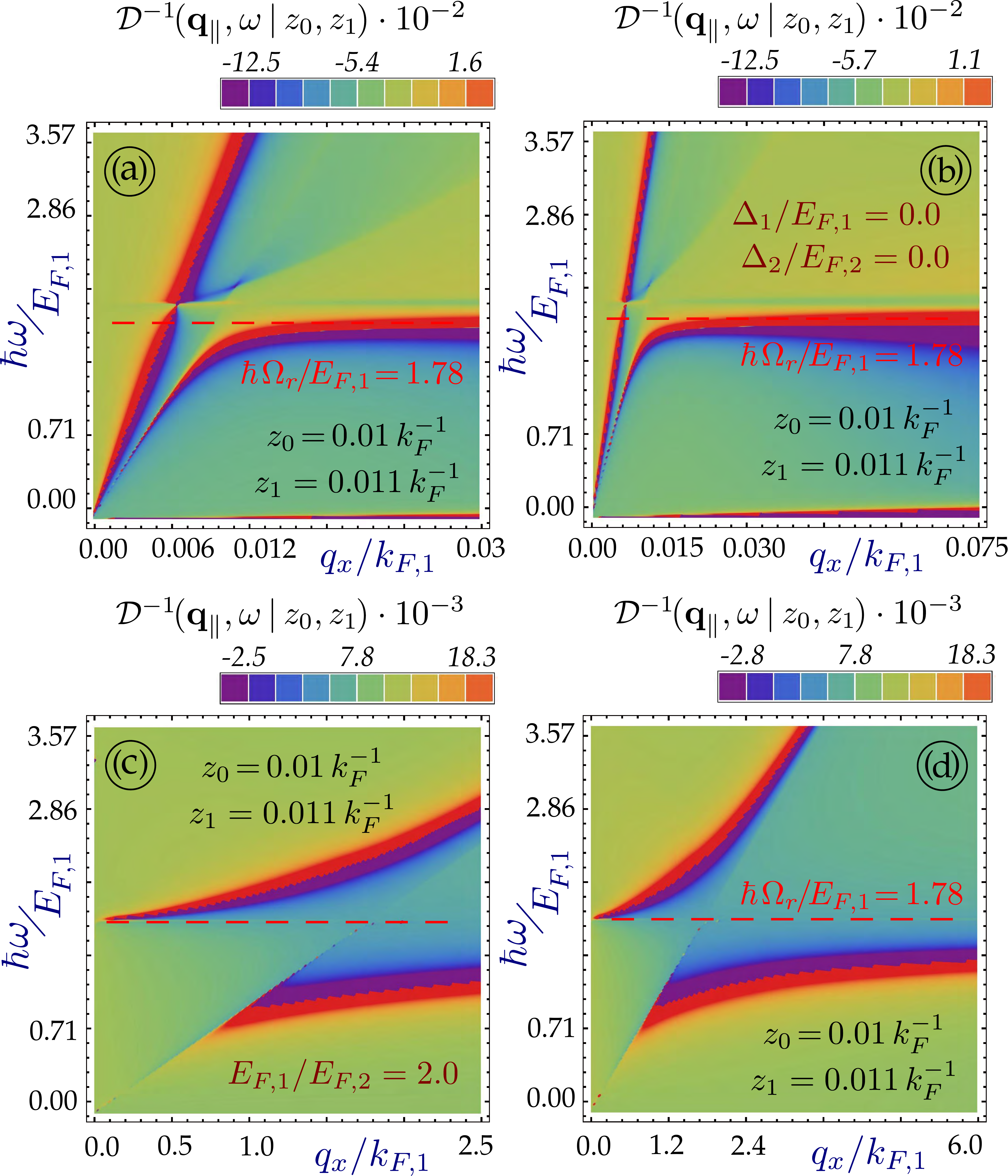}
\caption{(Color online)  Density plots for the real part of 
$\mathcal{D}^{-1}({\mbox{\boldmath$q$}}_\parallel, \omega \, \vert \, z_0, z_1)$
with $q_x^{\rm max}/k_{F} = 0.03,\,0.075,\,2.5$ and $6.0$. We chose $z_1=0.01\,k_{F}^{-1}$,
$z_2=0.011\,k_{F}^{-1}$ and $\hbar \Omega_r = 1.78\,E_{F}$.
Each graphene layer has zero bandgap, i.e.,  $\Delta_1=\Delta_2=0$ but has different 
Fermi energies $E_{F,1}=E_F$ and $E_{F,2}=2E_F$.}
\label{f5}
\end{figure}
In Fig.\,\ref{f5}, doping in the top layer is increased, thereby leading to two 
non-degenerate acoustic-like G-P modes, where the top layer has a higher G-P energy.
Compared with Fig.\,\ref{f3}, we find small but visible change in 
Figs.\,\ref{f5}($a$) and \ref{f5}($b$) for the optical-like SPP mode in 
shorter ranges for  $q_x$ since the graphene Fermi velocity is independent of doping.
For larger $q_x$ ranges in Figs.\,\ref{f5}($c$) and \ref{f5}($d$), the anticrossing 
gap is greatly increased due to  an enhanced retarded Coulomb interaction between the graphene 
layers and the conducting substrate for higher doping in the top layer. Meanwhile, 
the energy of the SPP mode is pushed up  significantly, which is attributed to the increased 
slope of the optical-like SPP mode by higher doping in the top layer.
However, the anticrossing is still dominated by the bottom layer G-P mode since the 
Landau damping of the top layer G-P mode becomes large due to its higher doping level.
as shown in Fig.\,\ref{f5}($d$).
\medskip

\begin{figure}
\centering
\includegraphics[width=0.48\textwidth]{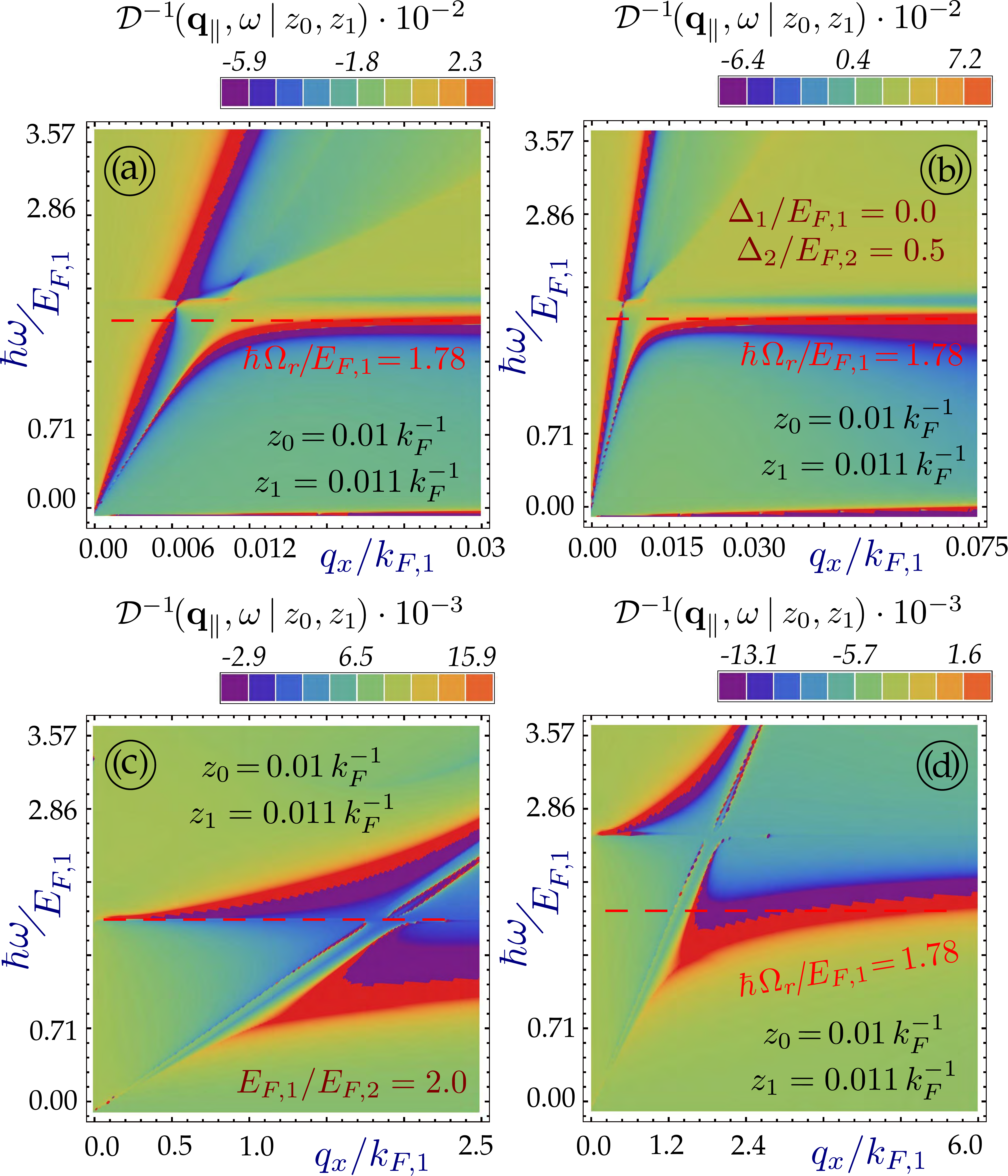}
\caption{(Color online)   Density plots for the real part of 
$\mathcal{D}^{-1}({\mbox{\boldmath$q$}}_\parallel, \omega \, \vert \, z_0, z_1)$
with $q_x^{\rm max}/k_{F} = 0.03,\,0.075,\,2.5$ and $6.0$. We chose $z_1=0.01\,k_{F}^{-1}$,
$z_2=0.011\,k_{F}^{-1}$ and $\hbar \Omega_r = 1.78\,E_{F}$.
Each graphene layer has a different Fermi energy given by 
 $E_{F,1}=E_F$ and $E_{F,2}=2E_F$ as well as
different bandgaps $\Delta_1=0$ and $\Delta_2=E_{F}$.}
\label{f6}
\end{figure}
The splitting of G-P modes in Fig.\,\ref{f4} with $\Delta_2=E_F/2$ becomes much more clear in Fig.\,\ref{f6} after we bring into a bandgap $\Delta_2=E_F$ to the top graphene layer in Fig.\,\ref{f5}.
We find no changes about photon and polariton modes in Figs.\,\ref{f6}($a$) and \ref{f6}($b$) for smaller $q_x$ ranges in comparison with Fig.\,\ref{f5}.
On the other hand, as the $q_x$ range is enlarged in Figs.\,\ref{f6}($c$) and \ref{f6}($d$), two non-degenerate acoustic-like G-P modes occur clearly in the anticrossing region, similar to Figs.\,\ref{f4}($c$) and \ref{f4}($d$).
Here, the increased doping level in the top layer pushes up the energy of optical-like SPP mode and expands the anticrossing gap, while the increased bandgap of the same layer splits the acoustic-like G-P mode into two
at the same time, in comparisons with Figs.\,\ref{f4} and \ref{f5}, respectively.
As a result, two successive plasmon-mode hybridizations can been see very clearly in Figs.\,\ref{f6}($c$) within the anticrossing region.
\medskip

\begin{figure}
\centering
\includegraphics[width=0.48\textwidth]{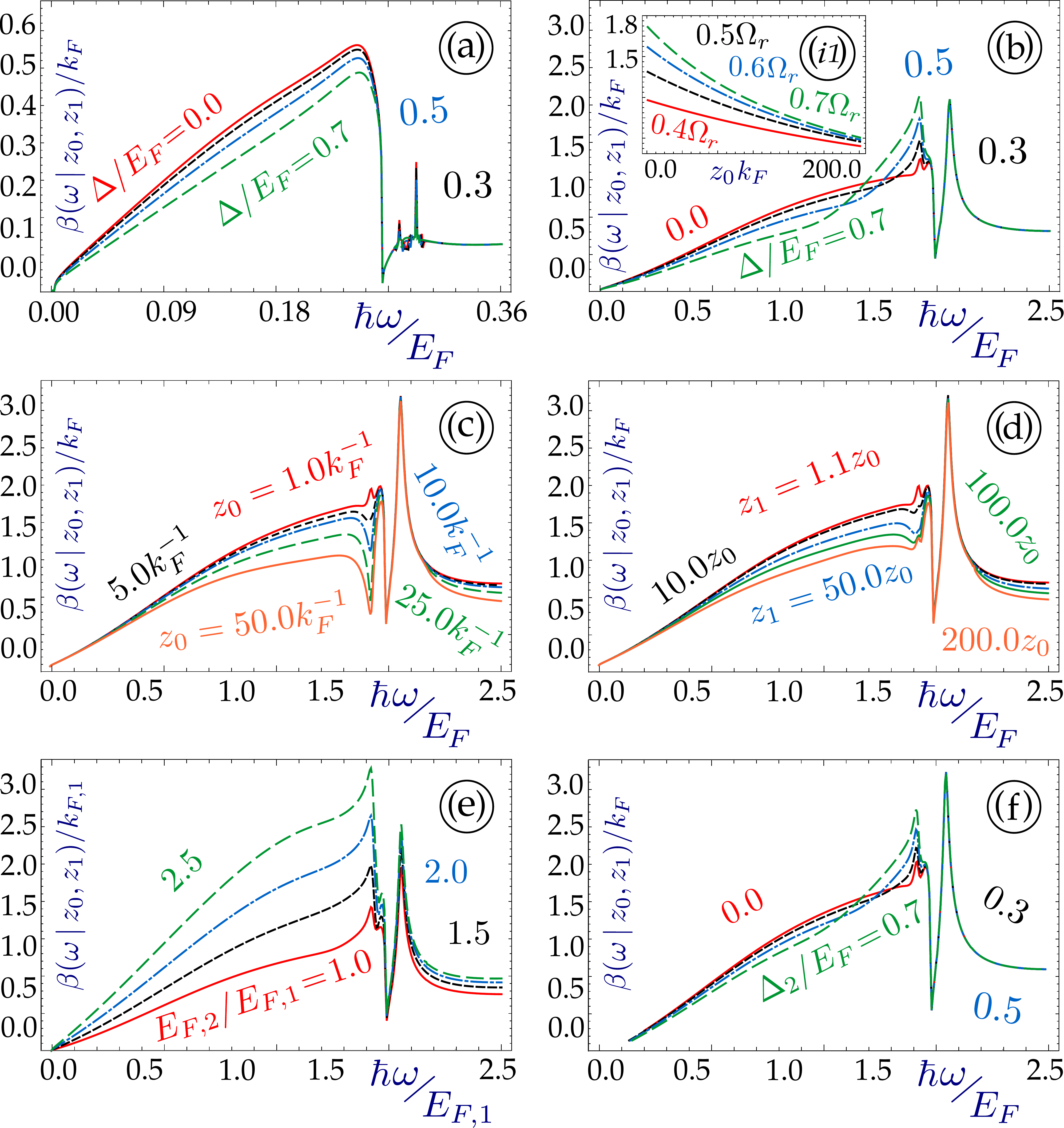}
\caption{(Color online)   Optical absorption spectra $\beta(\omega)$ (in units of 
$k_{F,1}$) as a function of scaled photon energy $\hbar\omega/E_{F,1}$.
Panels $(a)$ and $(b)$ are for two graphene layers at $z_0=0.01\,k^{-1}_{F}$, $z_1=1.1\,z_0$ 
and having $E_{F,1}=E_{F,2}=E_F$ (all but ($e$)), 
$\Delta_1=\Delta_2=\Delta$ (all but ($f$)) with
different SP energies $\hbar\Omega_r/E_F =0.25,\,1.78$, respectively.
The red, black, blue and green curves correspond to
$\Delta/E_{F}=0,\,0.3,\,0.5,\,0.7$, respectively.
Inset $(i1)$ shows the dependence of $\beta(\omega)$ on $z_0$ for
$\omega/\Omega_r= 0.4,\,0.5,\,0.6,\,0.7$.
All the other panels have $\hbar\Omega_r = 1.78\,E_{F,1}$.
Panel $(c)$ shows the red, black, blue, green and orange curves for 
$z_0/k^{-1}_{F}= 1,\,5,\,10,\,25,\,50$ with $z_1=2\,z_0$ and $\Delta=0$.
Plot $(d)$ displays the red, black, blue, green and orange curves for 
$z_1/z_0=1.1,\,10,\,50,\,100,\,200$ with $z_0=k^{-1}_{F}$ and $\Delta=0$.
Panel $(e)$ presents the red, black, blue and green curves for 
$E_{F,2}/E_{F,1}=1,\,1.5,\,2$ and 
$2.5$ with $\Delta/E_{F,1}=0.3$, $z_0 = 0.01\,k_{F,1}^{-1}$ and $z_2 = 1.1\,z_0$.
Plot $(f)$ displays the red, black, blue and green curves relate for 
$\Delta_2/E_F=0,\,0.3,\,0.5,\,0.7$ with $z_0 = 0.01\,k_{F}^{-1}$, $z_1 
= 1.1\,z_0$ and $\Delta_1=0$.}
\label{f7}
\end{figure}
The $z_j$ dependence in the secular equation 
${\cal D}et\left[\tilde{\cal C}^{jj'}_{\mu\nu}(q_x,\omega)\right]=0$ 
reflects the distinctive evanescent coupling between SPs and G-Ps.
By moving the graphene sheet a bit further away from the surface of the conductor  
(increasing $z_j$), the anticrossing gap will shrink due to
decreased retarded coupling between them. Meanwhile, the strengths of all 
the plasmon, polariton and photon modes will increase due to loss suppression 
of these modes to the conducting substrate. The incident SP field suffers not 
only Ohmic loss during its propagation along the conductor surface, but also 
absorption loss by its coupling to G-Ps.
\medskip

Figures\ \ref{f7}($a$) and \ref{f7}($b$) compare the absorption spectra 
$\beta_{\rm abs}(\omega)$ for $\hbar\Omega_r/E_F=0.25$ and $1.78$. When 
$\hbar\Omega_r$ is increased, the decay length of the SPP field becomes shorter,
as indicated by the inset of Fig.\,\ref{f7}($b$). Consequently, the SPP field 
will concentrate more within the region close to the conducting surface, and the 
overall absorptions of various plasmon modes look much stronger in 
Fig.\,\ref{f7}($b$). Here, the highest sharp peak in Fig.\,\ref{f7}($b$) is 
associated with the optical-like G-P mode which is hybridized with the SP mode.
The deep trough on its left-hand side results from the anticrossing gap.
Another peak with $\Delta\neq 0$ below this trough is attributed to the 
acoustic-like G-P mode which is accompanied by a hybrid SP peak on its right-hand 
side for $\Delta=0$ case. Finally, the rounded shoulder below this acoustic-like 
G-P peak comes from the SPP mode. As the bandgap $\Delta$ is increased, the hybrid 
SP peak is quickly suppressed, and  the acoustic-like G-P peak slightly moves 
down in energy from the trough side. Meanwhile, the SPP round peak is also 
reduced with increasing bandgap $\Delta$.
\medskip

When both graphene layers are moved further  away from the surface of the conductor
in Fig.\,\ref{f7}($c$),  there is little change in the highest optical-like 
G-P peak. On the other hand, the
acoustic-like G-P peak is completely suppressed for a larger layer separation from the surface, 
leading to a single sharp hybrid SP peak below the trough. This is further accompanied by 
the dramatic reduction of the SPP round peak. If only the interlayer separation is increased 
but the bottom layer is fixed, we find from Fig.\,\ref{f7}($d$) a very similar effect as 
that in Fig.\,\ref{f7}($c$). However, unresolved weak absorption from the bottom graphene 
layer still exists in this case.-
\medskip

In Fig.\,\ref{f7}($e$), we compare our results for double gapped-graphene layers having 
different doping levels in the top layer. The increased doping in the upper layer 
has no effect on the highest optical-like G-P peak and trough. Although the hybrid SP peak 
is suppressed by increasing the doping, the acoustic-like G-P peak is enhanced. More importantly, 
the rounded SPP peak increases greatly in this case by a large retarded Coulomb coupling to 
the conducting surface due to a higher doping level. If only the bandgap of the top layer is 
increased from zero, while that of the bottom layer is kept zero, we find a similar effect 
in Fig.\,\ref{f7}($f$) in comparison to that in Fig.\,\ref{f7}($b$), where the bandgaps of 
both layers are the same and increased from zero.
\medskip

\begin{figure}
\centering
\includegraphics[width=0.6\textwidth]{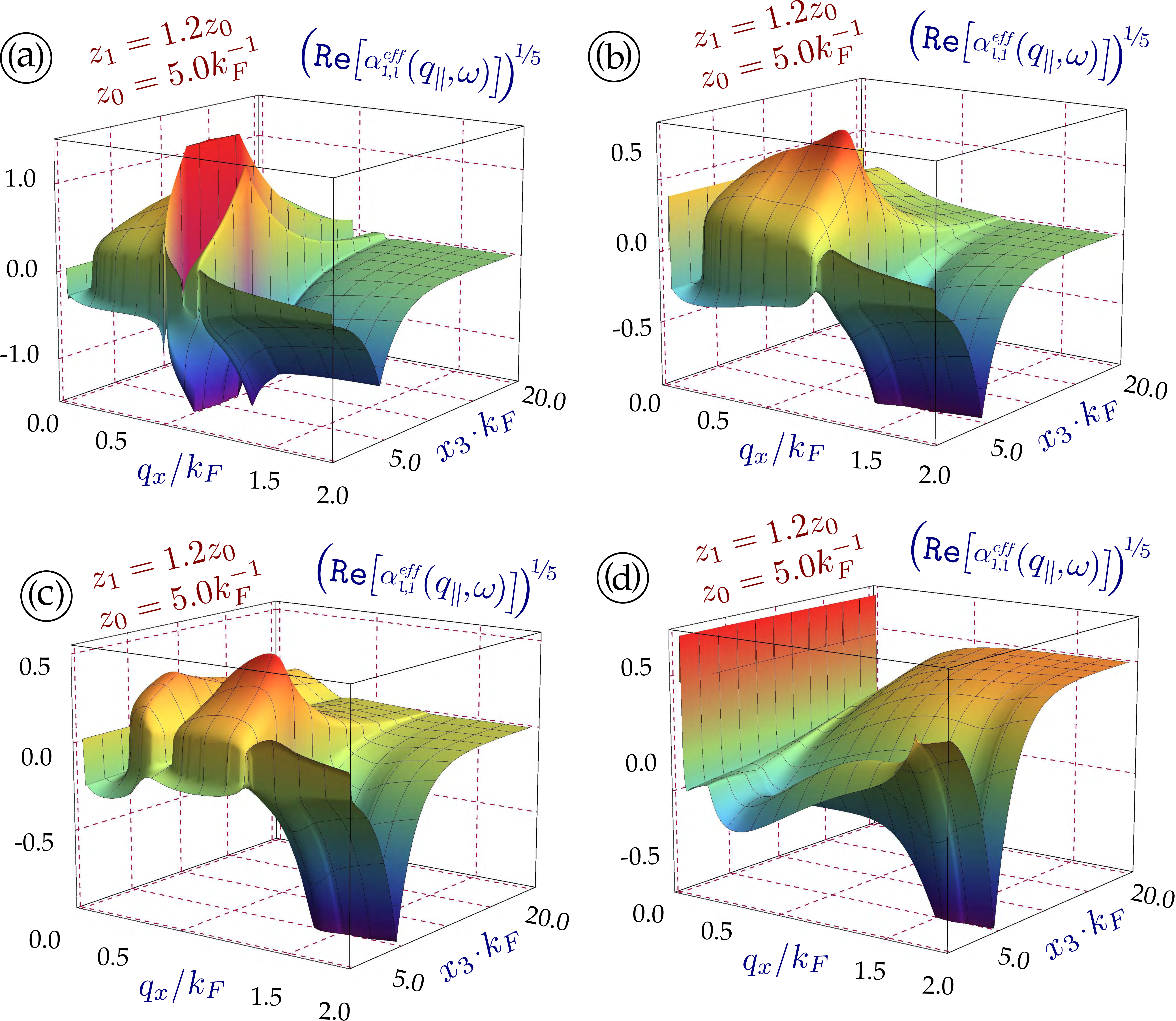}
\caption{(Color online)  3D plots for 
$[\text{Re}\{\alpha^{\rm eff}_{11} (q_x, \omega \, \vert \, x_3)\}]^{1/5}$
with $ \omega /\Omega_r= 0.7\, (a)$, $0.8\,(b)$, $0.9\,(c)$ and $1.0\,(d)$,
where $E_{F,1}=E_{F,2}=E_F$, $\Delta_1=\Delta_2=0$, $z_0 = 5.0\,k^{-1}_{F}$, $z_1 = 1.2\,z_0$, 
and $\hbar \Omega_r / E_F = 1.78$.}
\label{f8}
\end{figure}
In addition to  optical absorption by the G-Ps, SPs and SPPs in Fig.\,\ref{f7},   
resonant scattering of the SP from double-layer G-Ps also appears, as given by 
Eq.\,(\ref{eff-2}). Figure\ \ref{f8} presents 3D plots for
$[{\rm Re}\{\alpha^{\rm eff}_{11}(q_x,\omega\vert x_3)\}]^{1/5}$ with four 
$\omega$ values, where the two graphene sheets are placed relatively close to the 
surface. The scattering matrix is defined by $\alpha^{\rm eff}_{\mu\nu}\equiv\
delta(E_{\mu}-{\cal E}_{\mu}^{\rm inc})/\delta{\cal E}^{\rm inc}_{\nu}$, and therefore, 
its signs correspond to an enhanced ($+$) or weakened ($-$) SPP field after the scattering
with G-Ps. If both $q_x$ and $x_3$ are sufficiently large, such scattering is 
significantly  suppressed,  leaving only a wide and flat basin in the upper-right corners of 
Figs.\,\ref{f8}($a$)-\ref{f8}($d$). If $q_x$ is very small, the photon and SPP
 radiative modes dominate, and then, 
${\rm Re}\{\alpha^{\rm eff}_{11}(q_x,\omega\vert x_3)\}$ remains negative and 
becomes independent of $x_3$.When $q_x$ is intermediate, the SPP evanescent modes 
start entering in with increasing $\omega$ up to $\Omega_{r}$. 
In this case, the positive-peak strength is reduced and its peak coverage is squeezed 
into a smaller $x_3$ region where the localization of the SPP field is still insignificant.
In addition, the positive peak is broken into two islands in Fig.\,\ref{f8}($c$), and it 
switches to a negative peak followed by a negative constant in Fig.\,\ref{f8}($d$). 
On the other hand, when $q_x$ becomes very large for a strongly-localized SPP field,
its scattering by double-layer G-Ps becomes very small except for the resonance region 
very close to the surface as shown by the sharp negative edges in the lower-right corners 
of Figs.\,\ref{f8}($a$)-\ref{f8}($d$).
With increasing $\omega/\Omega_r$ in Fig.\,\ref{f7}($b$)-\ref{f7}($d$), it is 
interesting to note that this deep negative edge is pushed up to a large $q_x$ region 
due to SP resonance, and the $V$-shape feature at $q_x/k_F=2$
is sharpened simultaneously due to enhanced localization of the SPP field.
\medskip

\begin{figure}
\centering
\includegraphics[width=0.48\textwidth]{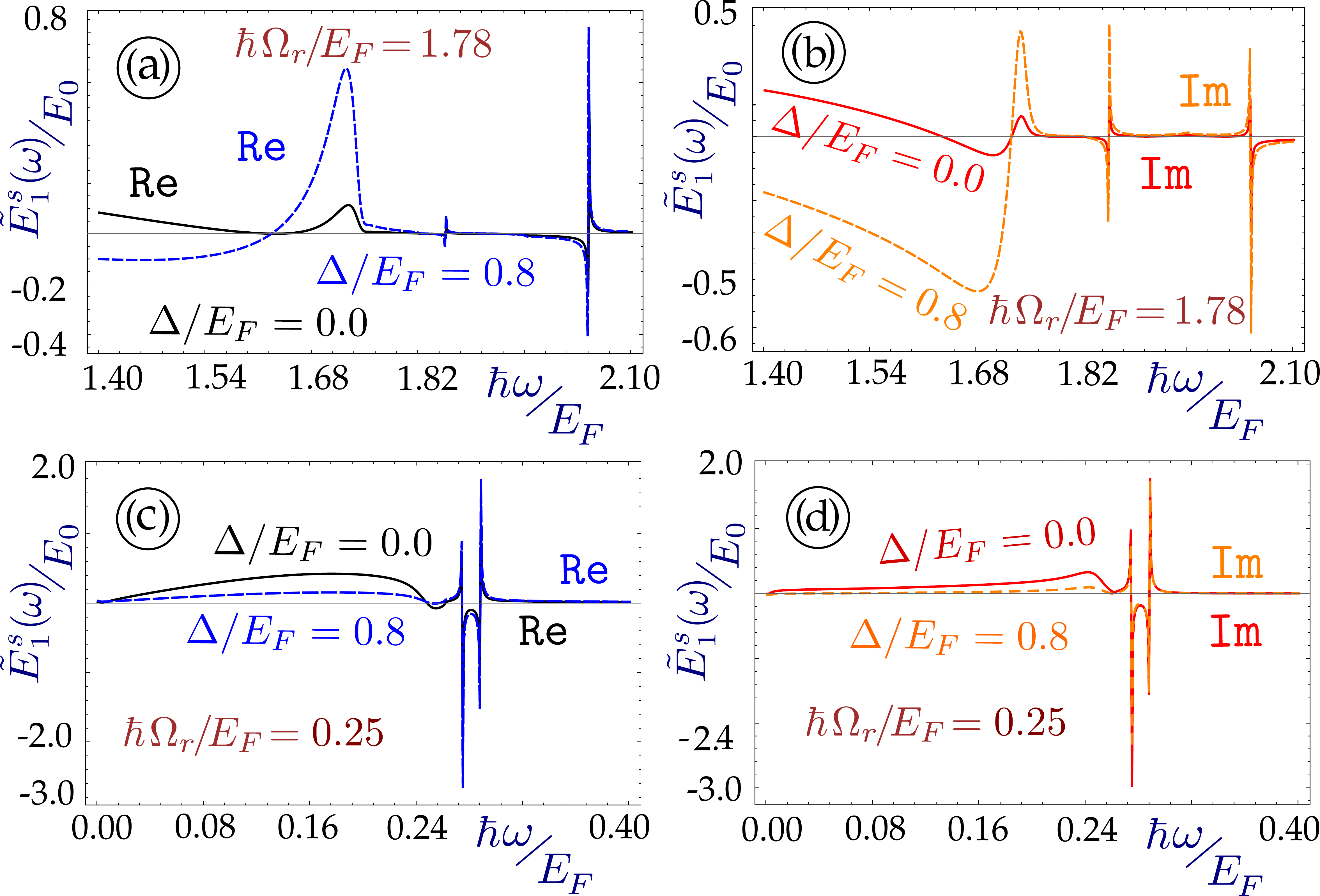}
\caption{(Color online)  Complex amplitudes $\tilde{E}_1^s(\omega\vert z_0)/E_0$ of 
a scattering field, calculated from Eq.\,(\ref{e17r}), as functions of incident 
photon energy $\hbar\omega/E_F$ for $x_3=z_0$, $\hbar\Omega_r/E_F= 1.78$ [($a$), ($b$)] 
and $\hbar\Omega_r/E_F= 0.25$ [($c$), ($d$)]. The real [($a$), ($c$)] and 
imaginary [($b$), ($d$)] parts of this complex amplitude are displayed 
for $\Delta/E_F=0$ and $0.8$ for each panel.  Here, we write the scattering field 
as: $\left.E^s_1(x_1,x_3\vert\omega)\right|_{x_3=z_0}\equiv\tilde{E}^s_1(\omega\vert z_0)\,\exp(i{\rm Re}[q_0(\omega)]x_1)$.
Moreover, we assume $E_{F,1}=E_{F,2}=E_F$, $\Delta_1=\Delta_2=\Delta$, $z_0 = 0.01\,k^{-1}_{F}$, and $z_1 = 200\,z_0$.}
\label{f9}
\end{figure}
In Figs.\,\ref{f9}($a$)-\ref{f9}($d$), we display the real and imaginary parts 
of calculated complex amplitudes $\tilde{E}_1^s(\omega\vert z_0)/E_0$ from Eq.\,(\ref{e17r}) 
as functions of $\hbar\omega/E_F$. When $\hbar\Omega_r/E_F=1.78$ in \ref{f9}($a$), we 
find a broad peak at $\omega=\Omega_r$ for the real part of $\tilde{E}_1^s(\omega\vert z_0)$ 
due to field scattering by the lower-energy SPP branch, which is further accompanied by a 
very (weak) strong plasmon resonance associated with field scattering by the higher-energy 
(acoustic-like) optical-like G-P branch. Similar peak and dual-plasmon-resonance features 
are also observed for the imaginary part of $\tilde{E}_1^s(\omega\vert z_0)$ in \ref{f9}($b$) 
but with an out-of-phase plasmon resonance for the optical-like G-Ps. 
Moreover, these unique scattering features in Figs.\,\ref{f9}($a$) and \ref{f9}($b$) 
are enhanced significantly with increased graphene bandgap $\Delta$. 
\medskip

Clearly, the peak and dual-plasmon-resonance features observed in 
Figs.\,\ref{f9}($a$) and \ref{f9}($b$) can be very well correlated to the 
absorption peaks in Fig.\,\ref{f7}. When the value of $\hbar\Omega_r/E_F$ is 
reduced from $1.78$ to $0.25$ in Figs.\,\ref{f9}($c$) and \ref{f9}($d$), on the 
other hand, the lowest broad SPP peak is greatly reduced for both real and 
imaginary parts of the complex amplitude $\tilde{E}_1^s(\omega\vert z_0)$. 
Meanwhile, the dual-plasmon-resonance (anticrossing-gap) region is shrunk dramatically, 
although the double scattering peaks by optical-like (right) and acoustic-like 
(left) G-P branches are still clearly visible. Furthermore, the increasing 
graphene bandgap $\Delta$ suppresses the SPP peak while it slightly enhances the 
dual-plasmon-resonance peaks at the same time. It is interesting to point out that the real 
and imaginary parts of the complex amplitude $\tilde{E}_1^s(\omega\vert z_0)$ in this 
case becomes in-phase in \ref{f9}($c$) and \ref{f9}($d$) for the 
right-most plasmon resonance associated with the optical-like G-P branch. 

\section{Conclusions and Remarks}
\label{sec-6}

The effect of electron back action on the hybridization of radiative and evanescent 
fields has been demonstrated by using a retarded interaction,
which is seen as hybrid dispersions for both radiative (small $q_x$ range) and 
evanescent (large $q_x$ range) field modes. Instead of a reaction force in Newtonian 
mechanics, the back action in this electro-optical study is an induced polarization 
field from the double-layer Dirac plasmons, which redistributes an incident 
surface-plasmon-polariton field by resonant scattering. The localization 
characteristics of such a retarded interaction ensures high sensitivity 
to dielectric environments surrounding and between the graphene sheets, 
including variations in the conducting substrate, cladding layer,
electronic properties of embedded graphene by a bandgap, as well as the graphene 
distance from the surface of the conductor. This provides a unique advantage in 
wavelength-selective optical scrutinizing for chemically-active molecules or 
proteins bounded with carbon atoms in graphene.
\medskip

The tools for optical probing which we discussed in this paper include either scattering 
or optical  absorption of an incident evanescent electromagnetic field. In the case of  
evanescent-field  scattering, we computed the spatial dependence of a Fourier transformed 
scattering matrix, which demonstrates the scattering enhancement,reduction and even suppression
as functions of graphene separations ($z_0$) from the surface of the conductor as well as between 
themselves ($z_1-z_0$) and the wave numbers ($q_x$) of the evanescent surface-plasmon-polariton 
field at several frequencies close to the localized surface-plasmon resonance.
This derived scattering matrix lays the foundation for constructing an effective-medium 
theory commonly employed in finite-difference time-domain methods\,\cite{add17,mit} for 
solving Maxwell's equations numerically. Furthermore, the calculated full spatial 
dependence for the scattering electromagnetic field shows unique features in three 
different regions, including ones below, between and above two graphene sheets.  
\medskip

For optical absorption, on the other hand, the triple peaks corresponding to the lower 
acoustic-like graphene plasmon, the middle surface-plasmon and the higher optical-like graphene 
plasmon modes  are seen to dominate the variable hybridization features at high conductor 
plasma frequencies. However, the rounded peak associated with the surface-plasmon-polariton 
mode at the lowest energy is found to be dominant at low plasma frequencies. In addition, 
this rounded peak further demonstrates that localized modes can be enhanced significantly 
when two graphene layers are placed closer to the conductor surface. These unique features in 
resonant absorption enable the selective excitation of radiative polariton modes for their 
condensation and a threshold-free laser afterwards.
\medskip

We would like to emphasize that the use of linear response theory\,\cite{ref2} 
for calculating the optical-response function in Eq.\,(\ref{e9}) only applies to a weak electromagnetic 
field. On the other hand, if the total electric field is strong, we must calculate the 
induced polarization field using the quantum-kinetic equations\,\cite{add8,add26,ref1,add9,add29}. 
In this case, the populations of electrons and holes in a density matrix become extreme 
non-equilibrium functions of wave vector for these photo-generated carriers. Moreover, the 
polarization field is determined by summing the light-induced coherence in the density 
matrix for all occupied states of photo-carriers. Furthermore, if the electric field is 
extremely strong, we expect an opening of energy gaps due to electron-photon dressing 
effects\,\cite{ref1,dressing,dressing2,kibis}. 
The theory for graphene-plasmon hybridization in this paper can be easily generalized to other 2D materials, such as silicene, germanene, molybdenum disulfide, etc. 

\begin{acknowledgments}
D.H. would like to thank the support from the Air Force Office of Scientific Research (AFOSR).
\end{acknowledgments}


\begin{references}
\bibitem{add8}H. Haug and S. W. Koch, \textsl{Quantum Theory of the Optical and Electronic Properties of Semiconductors} (Fourth Edition, World Scientific Publishing Co. Pte. Ltd., 2004).

\bibitem{add26}F. Rossi and T. Kuhn, Rev. Mod. Phys. {\bf 74}, 895 (2002).

\bibitem{ref1}D. H. Huang, M. M. Easter, G. Gumbs, A. A. Maradudin, S.-Y. Lin, D. A. Cardimona and X. Zhang, Opt. Expr. {\bf 22}, 27576 (2014).

\bibitem{add9}M. Lindberg and S. W. Koch, Phys. Rev. B {\bf 38}, 3342 (1988).

\bibitem{add29}M. Kira and S. W. Koch, Progress in Quantum Electronics {\bf 30}, 155 (2006).

\bibitem{new}A. Iurov, D. H. Huang, G. Gumbs, W. Pan, and A. A. Maradudin, Phys. Rev. B {\bf 96}, 081408(R) (2017).

\bibitem{add12}D. H. Huang, M. M. Easter, G. Gumbs, A. A. Maradudin, S.-Y. Lin, D. A. Cardimona, and X. Zhang, Appl. Phys. Lett. {\bf 104}, 251103 (2014).

\bibitem{swk}K. Schuh, M. Kolesik, E. M. Wright, J. V. Moloney, and S. W. Koch, Phys. Rev. Lett. {\bf 118}, 063901 (2017).

\bibitem{add1}K. Novoselov, A. K. Geim, S. Morozov, D. Jiang, M. Katsnelson, I. Grigorieva, S. Dubonos, and A. Firsov, Nature {\bf 438}, 197 (2005).

\bibitem{add2}A. K. Geim and K. S. Novoselov, Nature Materials {\bf 6}, 183 (2007).

\bibitem{add3}Y. Zhang, Y.-W. Tan, H. L. St\"ormer, and P. Kim, Nature {\bf 438}, 201 (2005).

\bibitem{add4}A. K. Geim, Science {\bf 324}, 1530 (2009).

\bibitem{special}G. Gumbs and D. H. Huang, ``\textsl{Electronic and Photonic Properties of Graphene Layers and Carbon Nanoribbons}'', Phil. Trans. R. Soc. A {\bf 368}, 5353 (2010).

\bibitem{chapter}G. Gumbs, D. H. Huang, A. Iurov, and B. Gao,``\textsl{Optoelectronic and transport properties of gapped graphene}'' in \textsl{Graphene Science Handbook: Electrical and Optical Properties} (Volume 3,
CRC Press, 2016) Chapter 30, pp. 489-504.

\bibitem{oe1}S. Christopoulos, G. B. H. von H\"ogersthal, A. J. D. Grundy, P. G. Lagoudakis, A.V. Kavokin, J. J. Baumberg, G. Christmann, R. Butt\'e, E. Feltin, J.-F. Carlin, and N. Grandjean, Phys. Rev. Lett. {\bf 98}, 126405 (2007).

\bibitem{oe2}S. I. Tsintzos, N. T. Pelekanos, G. Konstantinidis, Z. Hatzopoulos, and P. G. Savvidis, Nat. Lett. {\bf 453}, 372 (2008).

\bibitem{oe3}P. Bhattacharya, B. Xiao, A. Das, S. Bhowmick, and J. Heo, Phys. Rev. Lett. {\bf 110}, 206403 (2013).

\bibitem{oe4}C. Schneider, A. Rahimi-Iman, N. Y. Kim, J. Fischer, I. G. Savenko, M. Amthor, M. Lermer, A. Wolf, L. Worschech, V. D. Kulakovskii, I. A. Shelykh, M. Kamp, S. Reitzenstein, A. Forchel, Y. Yamamoto,
and S. H\"ofling, Nat. {\bf 497}, 348 (2013).

\bibitem{add28}E. L. Albuquerque and M. G. Cottam, Phys. Rep. {\bf 233}, 67 (1993).

\bibitem{ritchie}R. H. Ritchie, E. T. Arakawa, J. J. Cowan, and R. N. Hamm, Phys. Rev. Lett. {\bf 21}, 1530 (1968).

\bibitem{add5}B. Wang, X. Zhang, F. J. Garc\'ia-Vidal, X. Yuan, and J. Teng, Phys. Rev. Lett. {\bf 109}, 073901 (2012).

\bibitem{add6}M. Liu, X. Yin, and X. Zhang, Nano Letters {\bf 12}, 1482 (2012).

\bibitem{add7}M. Liu, X. Yin, E. Ulin-Avila, B. Geng, T. Zentgraf, L. Ju, F. Wang, and X. Zhang, Nature {\bf 474}, 64 (2011).

\bibitem{add10}F. Koppens, T. Mueller, P. Avouris, A. Ferrari, M. Vitiello, and M. Polini, Nature nanotechnology {\bf 9}, 780 (2014).

\bibitem{add11}A. A. Maradudin and D. L. Mills, Phys. Rev. B {\bf 11}, 1392 (1975).

\bibitem{add15}M. G. Cottam and A. A. Maradudin, ``Surface linear response functions'', in \textsl{Surface Excitations}, eds. V. M. Agranovich and R. Loudon (North-Holland, Amsterdam, 1984), pp. 1-194.

\bibitem{add27}J. M. Pitarke, V. M. Silkin, E. V. Chulkov, and P. M. Echenique, Rep. Prog. Phys. {\bf 70}, 1 (2007).

\bibitem{add24}A. V. Zayats, I. I. Smolyaninov, and A. A. Maradudin, Phys. Rep. {\bf 408}, 131 (2005).

\bibitem{pan2}G. Gumbs, A. Iurov, D. H. Huang, and W. Pan, J. Appl. Phys. {\bf 118}, 054303 (2015).

\bibitem{add14}A. Iurov, G. Gumbs, D. Huang, and V. Silkin, Phys. Rev. B {\bf 93}, 035404 (2016).

\bibitem{shawn}J. G. Fleming, S. Y. Lin, I. El-Kady, R. Biswas, and K. M. Ho, Nat. {\bf 417}, 52 (2002).

\bibitem{shawn2}B. J. Frey, P. Kuang, M.-L. Hsieh, J.-H. Jiang, S. John, and S.-Y. Lin, Sci. Rep. {\bf 7}, 4171 (2017).

\bibitem{add18}M. S. Tame, K. R. McEnery, S. K. \"Ozdemir, J. Lee, S. A. Maier, and M. S. Kim, Nat. Phys. {\bf 9}, 329 (2013).

\bibitem{add25}F. de Le\'on-P\'erez, G. Brucoli, F. J. García-Vidal, and L. Mart\'in-Moreno, New J. Phys. {\bf 10}, 105017 (2008).

\bibitem{add31}A. N. Grigorenko, M. Polini, and K. S. Novoselov, Nature Photonics {\bf 6}, 749 (2012).

\bibitem{add32}D. N. Basov, M. M. Fogler, A. Lanzara, F. Wang Y. Zhang, Rev. Mod. Phys. {\bf 86}, 959 (2014).

\bibitem{add19}D. S. Saxon, Phys. Rev. {\bf 100}, 1771 (1955).

\bibitem{add30}F. J. Garc\'ia de Abajo, Rev. Mod. Phys. {\bf 79}, 1267 (2007).

\bibitem{add20}J. van Kranendonk and J. E. Sipe, ``Foundations of the macroscopic electromagnetic theory of dielectric media'', in \textsl{Progress in Optics XV}, ed. E. Wolf (New York: North-Holland, 1977), Chap. 5.

\bibitem{add21}G. D. Mahan and G. Obermair, Phys. Rev. {\bf 183}, 834 (1969).

\bibitem{add22}J. Sipe and J. van Kranendonk, Phys. Rev. A {\bf 9}, 1806 (1974).

\bibitem{add23}W. Lamb, D. M. Wood, N. W. Ashcroft, Phys. Rev. B {\bf 21}, 2248 (1980).

\bibitem{xzhang}X. Zhang and Z. Liu, Nat. Mater. {\bf 7}, 435 (2008).

\bibitem{add13}M. Wojcik, M. Hauser, W. Li, S. Moon, and K. Xu, Nature Communications {\bf 6}, 7384 (2015).

\bibitem{ieee}D. H. Huang, P. M. Alsing, D. A. Cardimona, and G. Gumbs, IEEE Trans. Nanotechn. {\bf 7}, 151 (2008).

\bibitem{spie}D. H. Huang, O. Roslyak, G. Gumbs, W. Pan and A. A. Maradudin, Proc. SPIE {\bf 9961}, 996104 (2016).

\bibitem{ref2}G. Gumbs and D. H. Huang, \textsl{Properties of Interacting Low-Dimensional Systems} (John Wiley \& Sons, 2011), Chap. 2.

\bibitem{ref2-1}G. Gumbs and D. H. Huang, \textsl{Properties of Interacting Low-Dimensional Systems} (John Wiley \& Sons, 2011), Chap. 4.

\bibitem{ref3}D. H. Huang, G. Gumbs and O. Roslyak, Appl. Opt. {\bf 52}, 755 (2013).

\bibitem{add16}O. Roslyak, G. Gumbs and D. H. Huang, J. Appl. Phys. {\bf 109}, 113721 (2011).

\bibitem{zero}B. Wunsch, T. Stauber, F. Sols, and F. Guinea, New J. Phys. {\bf 8}, 318 (2006).

\bibitem{sp}R. H. Ritchie, Phys. Rev. {\bf 106}, 874 (1957).

\bibitem{new-1}A. Principi, M. Polini, and G. Vignale, Phys. Rev. B {\bf 80}, 075418 (2009).

\bibitem{new-2}T. Stauber and G. G\'omez-Santos, Phys. Rev. B {\bf 82}, 155412 (2010).

\bibitem{abs}G. Gumbs, D. H. Huang, and D. N. Talwar, Phys. Rev. B {\bf 53}, 15436 (1996).

\bibitem{lorentz}D. H. Huang and Y. Zhao, Phys. Rev. A {\bf 51}, 1617 (1995).

\bibitem{abs2}G. Gumbs and D. H. Huang, Phys. Rev. B {\bf 50}, 15148 (1994).

\bibitem{abs3}D. H. Huang, G. Gumbs, and N. J. M. Horing, Phys. Rev. B {\bf 49}, 11463 (1994).

\bibitem{add17}A. F. Oskooi, D. Roundy, M. Ibanescu, P. Bermel, J. D. Joannopoulos, S. G. Johnson, Computer Physics Communications {\bf 181}, 687 (2010).

\bibitem{mit}J. D. Joannopoulos, S. G. Johnson, J. N. Winn, and R. D. Meade, \textsl{Photonic Crystals: Molding the Flow of Light}  (Second Edition, Princeton University Press, 2008). 

\bibitem{dressing}A. Iurov, G. Gumbs, O. Roslyak, and D. H. Huang, J. Phys.: Condens. Matter {\bf 25}, 135502 (2013).

\bibitem{dressing2}A. Iurov, L. Zhemchuzhna, G. Gumbs, and D. H. Huang, J. Appl. Phys. {\bf 122}, 124301 (2017).

\bibitem{kibis}O. V. Kibis, Phys. Rev. B {\bf 81} 165433 (2010).
\end{references}
\end{document}